\documentclass[traditabstract]{aa}
\RequirePackage{graphicx,natbib,txfonts,rotating,lscape}

\begin{document}

\titlerunning{A Radiative Transfer Model of IRDCs in the Hi-GAL Survey}
\title{The initial conditions of high-mass star formation: radiative transfer models of IRDCs seen in 
the \textit{Herschel}\thanks{\textit{Herschel} is an ESA space observatory with science instruments provided by
European-led Principal Investigator consortia and with important participation from NASA.} Hi-GAL survey}
\author{L. A. Wilcock\inst{1} \and J. M. Kirk\inst{1} \and D. Stamatellos\inst{1} \and D. Ward-Thompson\inst{1} \and A.
Whitworth\inst{1}\and C. Battersby\inst{2} \and C. Brunt\inst{3} \and G. A. Fuller\inst{4} \and M. Griffin\inst{1}
\and S. Molinari\inst{5} \and P. Martin\inst{6} \and J. C. Mottram\inst{3} 
\and N. Peretto\inst{4}$^,$\inst{7} \and R. Plume\inst{8} \and H. A. Smith\inst{9} \and M. A. Thompson\inst{10}} 

\institute{School of Physics and Astronomy, Cardiff University, Queen's Building, Cardiff, CF24 3AA, UK 
\and Center for Astrophysics \& Space Astronomy, University of Colorado, Boulder, Colorado, 80309, USA 
\and School of Physics, University of Exeter, Stocker Road, Exeter, EX4 4QL, UK 
\and Jodrell Bank Centre for Astrophysics, School of Physics and Astronomy, University of Manchester, Manchester, M13 9PL, UK 
\and Instituto di Fisica dello Spazio Interplanetario, CNR, via Fosso del Cavaliere, I-00133 Roma, Italy
\and Canadian Institute for Theoretical Astrophysics, University of Toronto, Toronto, Canada, M5S 3H8
\and  Laboratoire AIM, CEA/DSM-CNRS-Universit\'e Paris Diderot, IFRU/Service d'Astrophysique, C.E. Saclay, 
Orme des merisiers, 91191 Gif-sur-Yvette, France 
\and University of Calgary, Dept Physics-Astronomy, Calgary, AB T2N 1N4, Canada 
\and Harvard-Smithsonian Center for Astrophysics, 60 Garden Street, Cambridge, MA, 02138, USA
\and Centre for Astrophysics Research, Science and Technology Research Institute, University of Hertfordshire, AL10 9AB, UK}

\date{Recieved 28/07/2010 / Accepted 07/12/2010}

\abstract{The densest infrared dark clouds (IRDCs) may represent the earliest observable stage of high-mass star 
formation. These clouds are very cold, hence they emit mainly at far-infrared and sub-mm wavelengths. For the first time,
\textit{Herschel} has
provided multi-wavelength, spatially resolved observations of cores within IRDCs, which, when combined with 
radiative transfer modelling, can constrain their properties, such as mass, density profile and dust temperature. We use a 3D,
multi-wavelength Monte Carlo radiative transfer code to model in detail the emission from six cores in three typical 
IRDCs seen in the Hi-GAL survey (G030.50+00.95, G031.03+00.26 and G031.03+00.76),
 and thereby to determine the properties of these cores and compare them with their low-mass equivalents. We found
 masses ranging from 90 to 290\,M$_{\odot}$ with temperatures from 8 to 11\,K at the centre of each core and 18 to 28\,K at the
 surface. The maximum luminosity of an embedded star within each core was calculated, and we rule out the possibility of
 significant high mass star formation having yet occurred in three of our cores. }


\maketitle

\begin{table*}
\begin{center}
\caption{The physical properties of the six cores.}
\label{coreprop}
\begin{tabular}{lllllll} \hline
 & \multicolumn{2}{l}{G030.50+00.95} & \multicolumn{2}{l}{G031.03+00.26} & \multicolumn{2}{l}{G031.03+00.76} \\
 & Core A & Core B & Core A & Core B & Core A & Core B \\ \hline 
R.A. (2000) & 18:43:34.7 & 18:43:34.9 & 18:47:01.5  & 18:47:04.2 & 18:45:10.2 & 18:45:17.3 \\
Dec. (2000) & $-$01:42:29 & $-$01:44:36 & $-$01:34:36  & $-$01:33:39 & $-$01:20:00 & $-$01:19:12 \\
Name in PF09 & \multicolumn{2}{l}{SDC30.442+0.958} & \multicolumn{2}{l}{SDC31.039+0.241} & \multicolumn{2}{l}{Not included} \\
Distance (pc) & 3400 & 3400 & 4900 & 4900 & 3400 & 3400 \\
Semi-major Axis (pc) & 0.6 & 0.7 & 0.45 & 0.55 & 0.4 & 1.0 \\
\textit{R$_0$} (pc) & 0.06 & 0.07 & 0.04  & 0.05 & 0.04 & 0.1 \\
Aspect ratio & 1.93 & 3.16 & 2.46 & 2.22 & 1.70 & 1.65 \\
Position Angle & 50 & 85 & 125 & 340 & 335 & 85 \\
\textit{n}$_{0}$(H$_2$) (cm$^{-3}$) & 1.4$\times$10$^5$ & 4.5$\times$10$^4$ & 3.8$\times$10$^5$ & 4.2$\times$10$^5$ & 6.0$\times$10$^5$ &
2.5$\times$10$^4$ \\
ISRF$_{Habing}$ & 8.5 & 18 & 18 & 81 & 17 & 4.8 \\
$\tau_{\theta=90\degr} / \tau_{\theta=0\degr}$ & 1.93 & 3.16 & 2.46 & 2.22 & 1.59 & 1.65 \\
$F_{500\, \mu m}$ (Jy) & 5.0$\pm0.5$ & 9.0$\pm0.9$ & 3.3$\pm0.3$ & 10$\pm1$ & 4.5$\pm0.5$ & 4.4$\pm0.4$ \\
Model Temperature (K) & 9--19 & 11--22 & 9--22 & 11--28 & 8--22 & 10--18 \\
SED Temperature (K) & 14$\pm1$ & 16$\pm2$ & 14$\pm1$ & 17$\pm2$ & 14$\pm1$ & 14$\pm1$ \\
Model mass (M$_\odot$) & 120 & 140 & 170 & 290 & 110 & 90 \\
SED mass -- $\kappa$=0.05\,cm$^{2}$\,g$^{-1}$ (M$_\odot$) & 130 & 170 & 170 & 310 & 110 & 100 \\
SED mass -- $\kappa$=0.03\,cm$^{2}$\,g$^{-1}$ (M$_\odot$) & 210 & 280 & 280 & 510 & 180 & 160 \\ 
$ln\left(\frac{I_{1}}{I_{2}}\right)$ & 0.75 & -- & 1.50 & -- & 1.25 & -- \\
$L_{max}$ (\textit{L$_{\odot}$}) & $<$6 & $<$3 & $<$200 & $<$1000 & $<$40 & $<$9 \\
\hline
\end{tabular}
\tablefoot{
\textit{R$_0$} is the flattening radius within the core, 
\textit{n}$_{0}$(H$_2$) is the central density of each core,   
ISRF$_{Habing}$ is the ISRF of the core at FUV given in multiples of the Habing flux --- i.e. 1.6$\times$10$^{-3}$\, 
erg~s$^{-1}$~cm$^{-2}$ \protect\citep{habing68}, 
$\tau_{\theta=90\degr} / \tau_{\theta=0\degr}$ is the ratio of optical depths at $\theta$=90\degr and at $\theta$=0\degr and  
$F_{500\, \mu m}$ is the flux density of each core at 500\,$\mu$m. 
The model temperature is the range of temperatures in the core calculated from the model, and 
the SED temperature was calculated from single-temperature greybody fitting.
The model mass was calculated using $\kappa$=0.05\,cm$^2$g$^{-1}$ at 500\,$\mu$m (\protect\citealp{ossenkopf94}). 
The two SED masses were calculated using the method of \protect\citet{hildebrand83}, the SED temperature and the flux density at 
              500\,$\mu$m (see Section \ref{mass}). 
The difference in ISRFs, $ln\left(\frac{I_{1}}{I_{2}}\right)$, was calculated from the model (see Section \ref{DRF}).
$L_{max}$ is the maximum luminosity of a star that could be embedded in each core (see Section \ref{heat}).}
\end{center}
\end{table*}

\section{Introduction}\label{intro}
We present observations, performed with the ESA \textit{Herschel} Space Observatory \citep{herschel}, of a section of the
Galactic Plane. We use \textit{Herschel}'s large collecting area and powerful science payload to perform imaging photometry
using the PACS (Photodetector and Array Camera Spectrometer; \citealt{pacs}) and SPIRE (Spectral and Photometric Imaging Receiver;
\citealt{spire}) instruments. These observations were carried out as part of \textit{Herschel}'s Science Demonstration Phase (SDP) 
for the \textit{Herschel} Infrared Galactic Plane Survey (Hi-GAL; \citealt{higalb, higala}) open time Key Project to map the inner 
Galactic Plane. In particular, we are studying cores within the infrared dark clouds (IRDCs) found in the Galactic Plane. 

IRDCs were intially discovered by the ISO \citep{perault96} and MSX \citep{carey98, egan98}
surveys as regions of high contrast against the mid-infrared (MIR) background. The densest IRDCs may eventually form massive stars,
and are therefore presumed to represent the earliest observable stage of high mass star formation. Some IRDCs contain 
cold, compact cores, which are believed to be the high mass equivalent of low-mass prestellar cores. 

IRDCs and their cores have low temperatures (T$<$25\,K; e.g. \citealt{egan98, teyssier02}) and therefore emit mostly at
far-infrared (FIR) and sub-mm wavelengths. However, until now, spatially resolved observations have not been
available in the FIR. \textit{Herschel} changes this, and offers high resolution FIR images of not only
IRDCs, but of the individual cores within -- potentially allowing, for the first time, temperatures to be determined and column 
densities and masses to be accurately constrained for a large number of IRDCs. 
In this paper we present some first examples of how the \textit{Herschel} data can be used to trace the physical parameters 
of cores within IRDCs.

\section{A Sample of IRDC Cores in the Hi-GAL Science Demonstration Fields}

Hi-GAL is an Open Time Key Project of the \textit{Herschel} Space Observatory and is performing a survey of the inner Galactic Plane 
($|$\textit{l}$|$$<$60\degr, $|$\textit{b}$|$$<$1\degr) using the PACS and SPIRE instruments. The two are used in parallel mode to 
map the Milky Way Galaxy simultaneously at five wavelengths, from 70 to 500\,$\mu$m, with resolutions up to 5\arcsec{} at 70\,$\mu$m. 

Two regions of the Hi-GAL survey were completed during SDP time. The regions are each four degrees square and center
around \textit{l}=30\degr, \textit{b}=0\degr and \textit{l}=59\degr, \textit{b}=0\degr. Using these observations and
the \citet[hereafter S06]{simon06c} and \citet[hereafter PF09]{peretto09} catalogues, IRDCs can be identified in the Galactic Plane 
with none of the uncertainty that exists in using MIR observations alone (e.g. being unable to differentiate between an IRDC and a 
dip in the MIR background). 

PACS data reduction was performed using the \textit{Herschel} Interactive Pipeline Environment (HIPE; \citealt{hipe}), albeit with some
departures from the standard processing \citep{pacs}. Most notably, the standard deglitching and crosstalk correction were not used 
due to poor results in these fields, and custom procedures were written for drift removal. 

SPIRE data processing required less deviation
from the standard processing methods \citep{spire}, with both standard deglitching and drift removal producing acceptable results. 
In both cases, the ROMAGAL Generalised Least Squares algorithm \citep{traficante10} was used to produce the final maps. A more 
thorough discussion of the entire data reduction process can be found in \citet{traficante10}.

Each of the S06 sources in these regions was studied at 6 wavebands, i.e. \textit{Spitzer} GLIMPSE 8\,$\mu$m \citep{glimpse, spitzer} 
and the five Hi-GAL wavelengths (70, 160, 250, 350 and 500\,$\mu$m). If the source showed extinction at
MIR but emission in the FIR it was identified as an IRDC. Within the \textit{l}=30\degr{} region, approximately 330 
(PF09) IRDCs were found, each containing at least one infrared dark core. 

\begin{sidewaysfigure*}
\begin{center}
\includegraphics[angle=0,width=40mm]{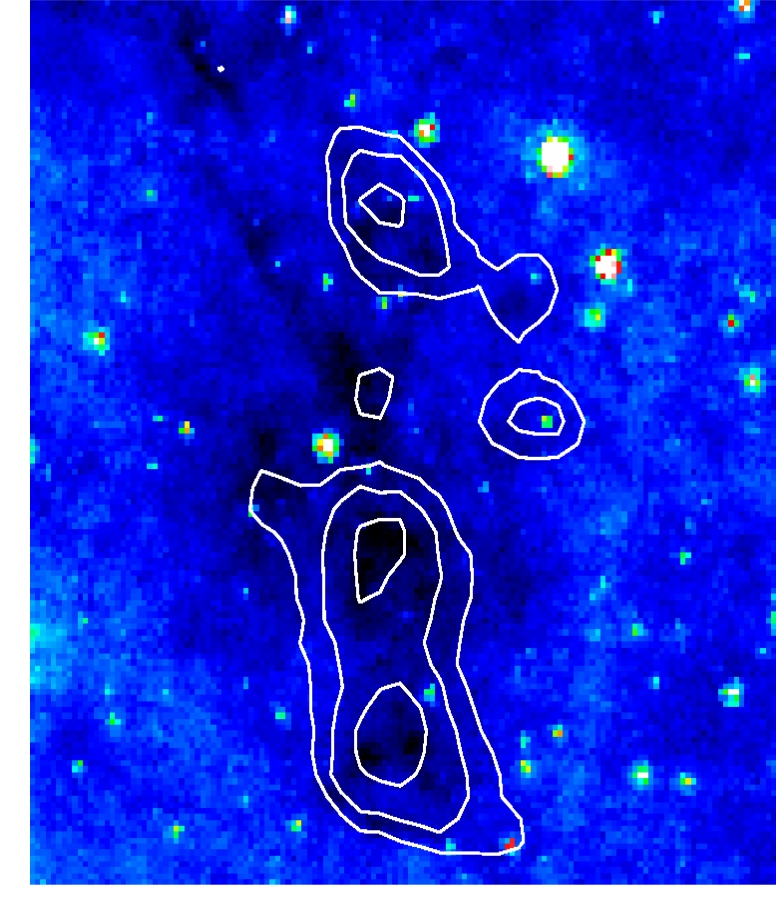}
\includegraphics[angle=0,width=40mm]{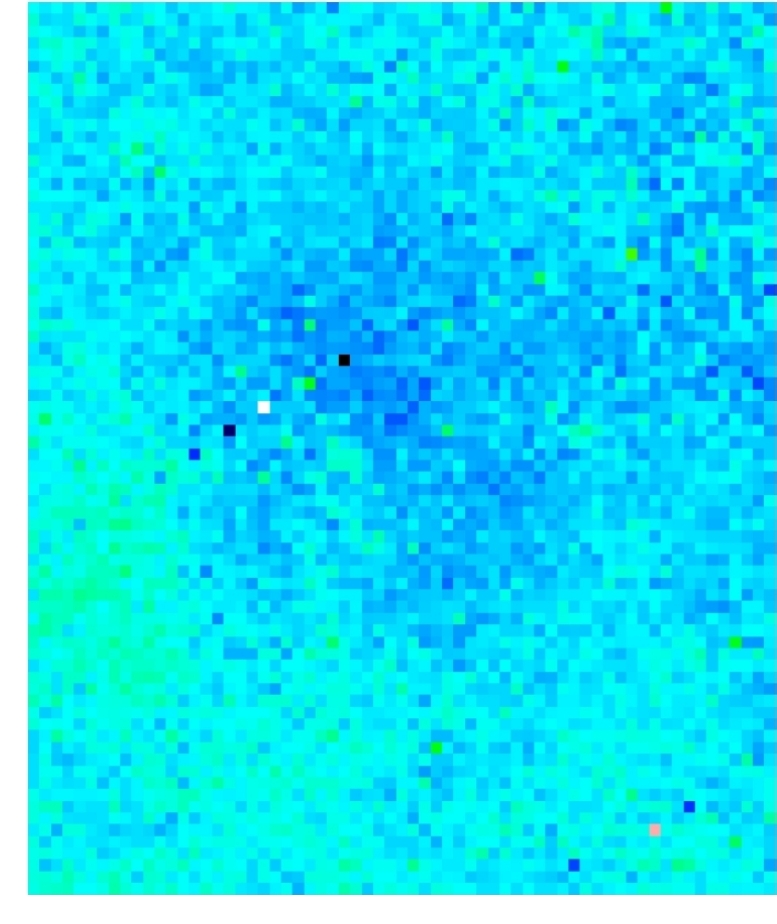}
\includegraphics[angle=0,width=40mm]{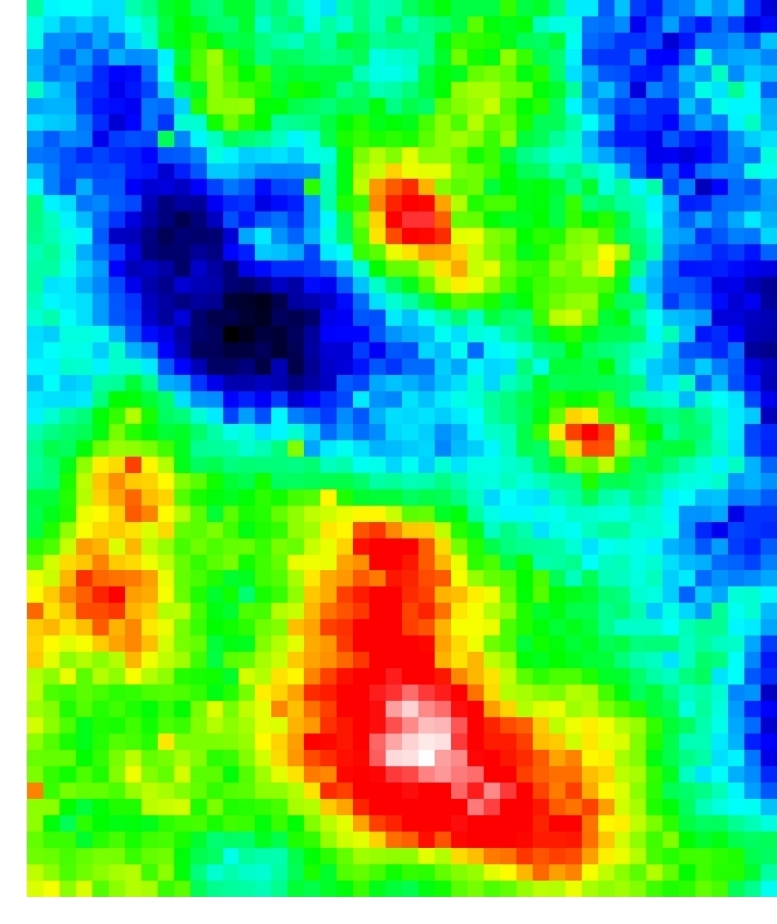}
\includegraphics[angle=0,width=40mm]{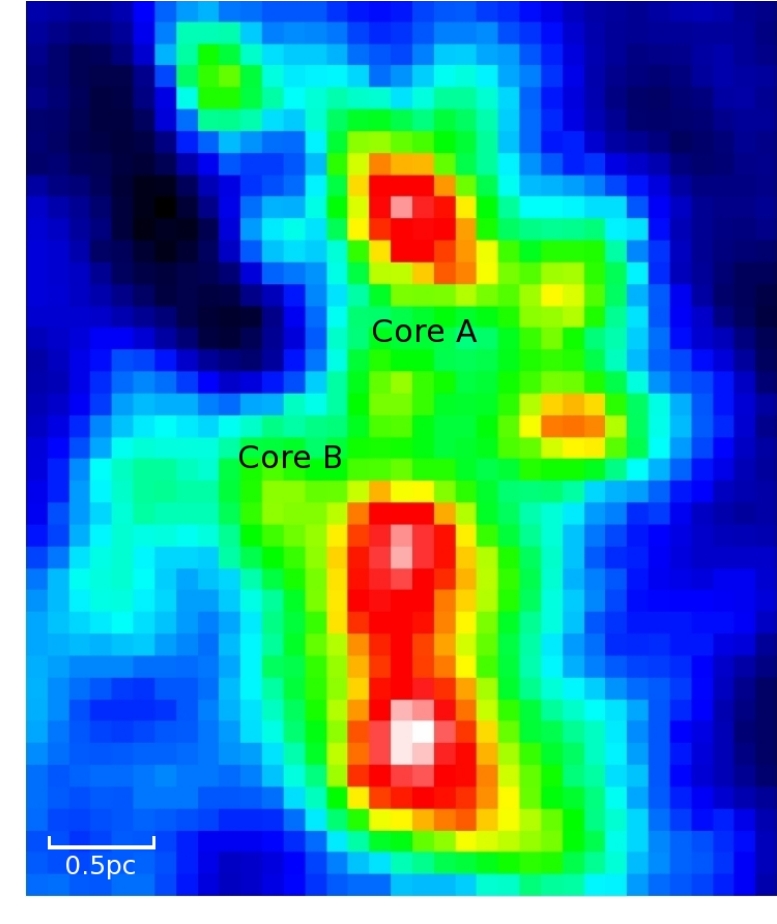}
\includegraphics[angle=0,width=40mm]{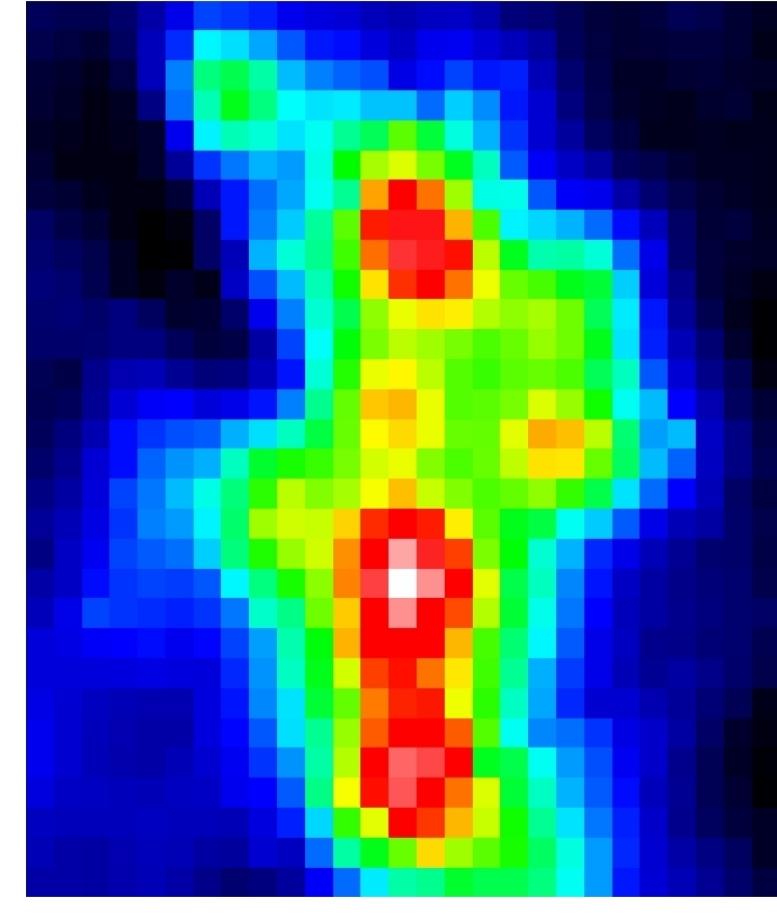}
\includegraphics[angle=0,width=40mm]{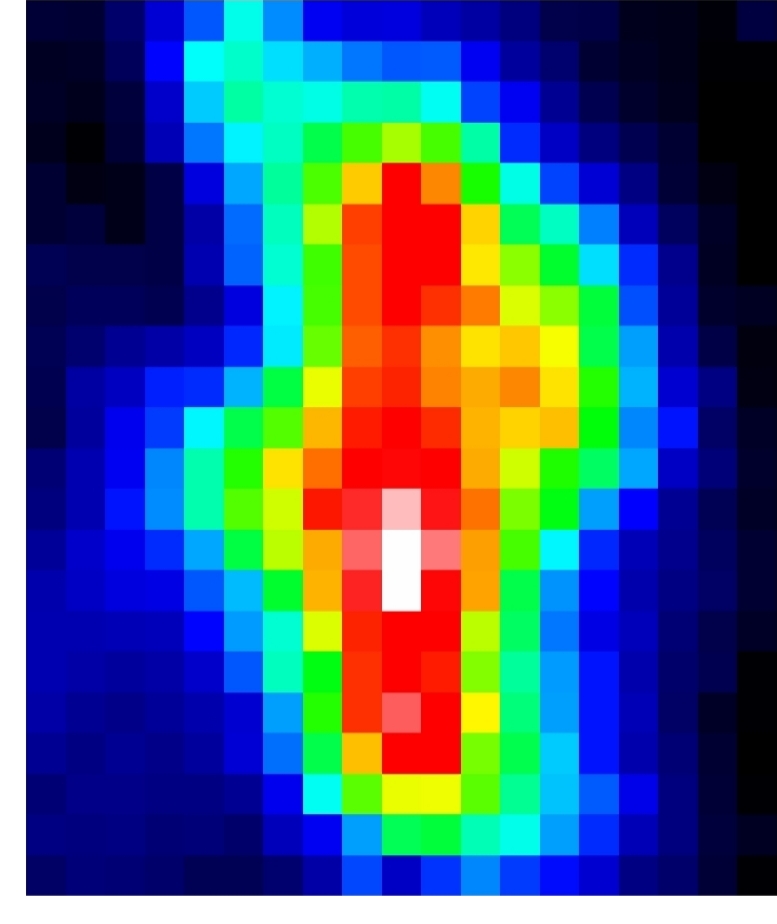}
\includegraphics[angle=0,width=40mm]{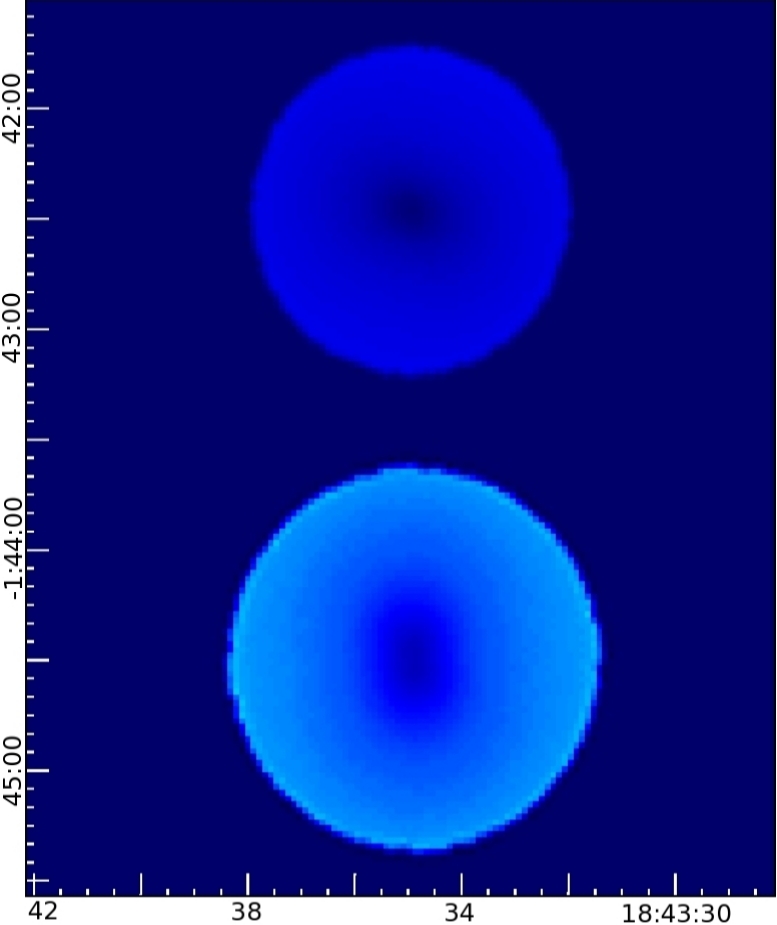}
\includegraphics[angle=0,width=40mm]{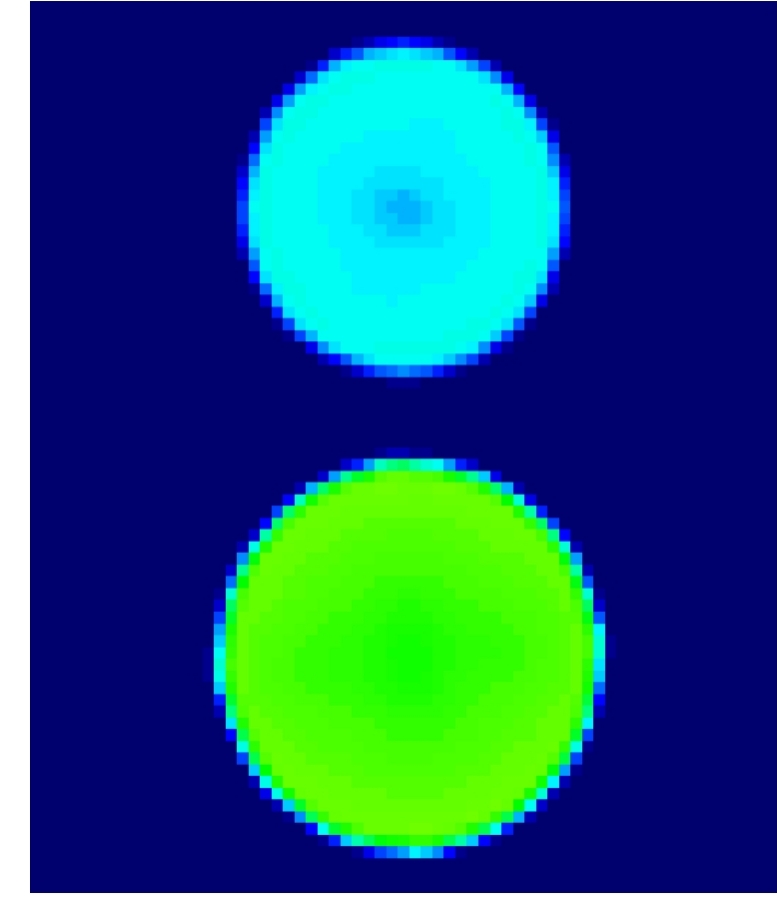}
\includegraphics[angle=0,width=40mm]{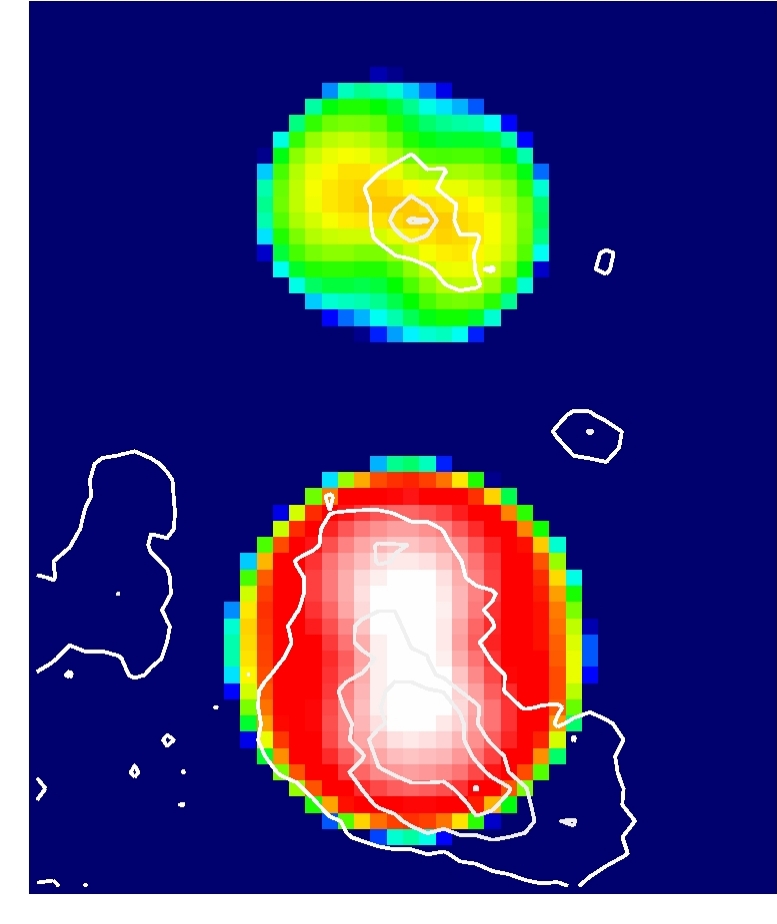}
\includegraphics[angle=0,width=40mm]{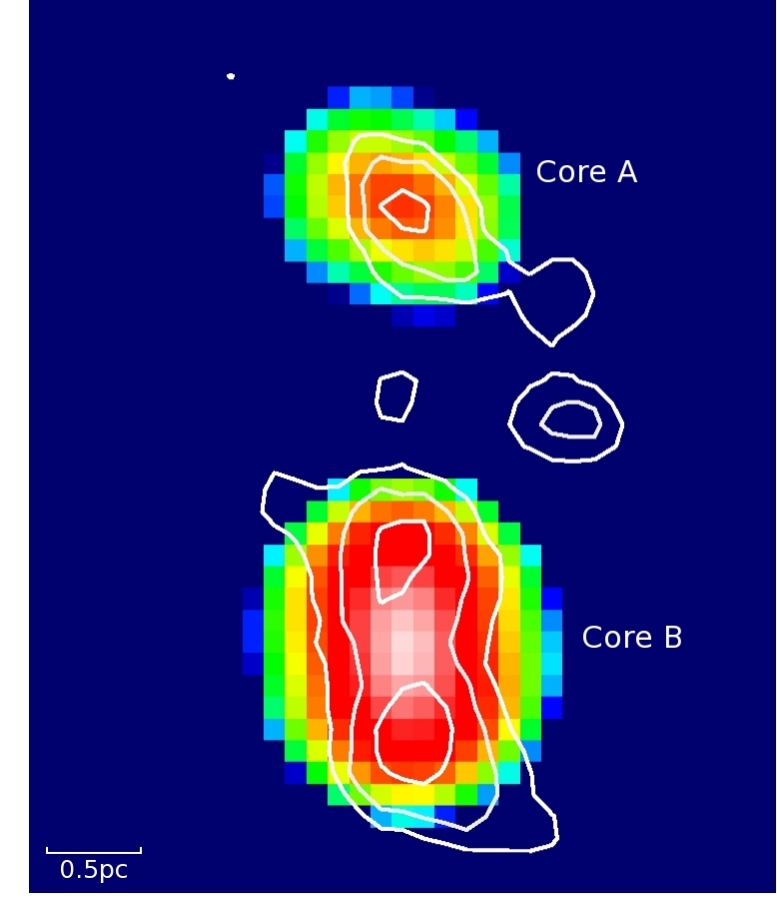}
\includegraphics[angle=0,width=40mm]{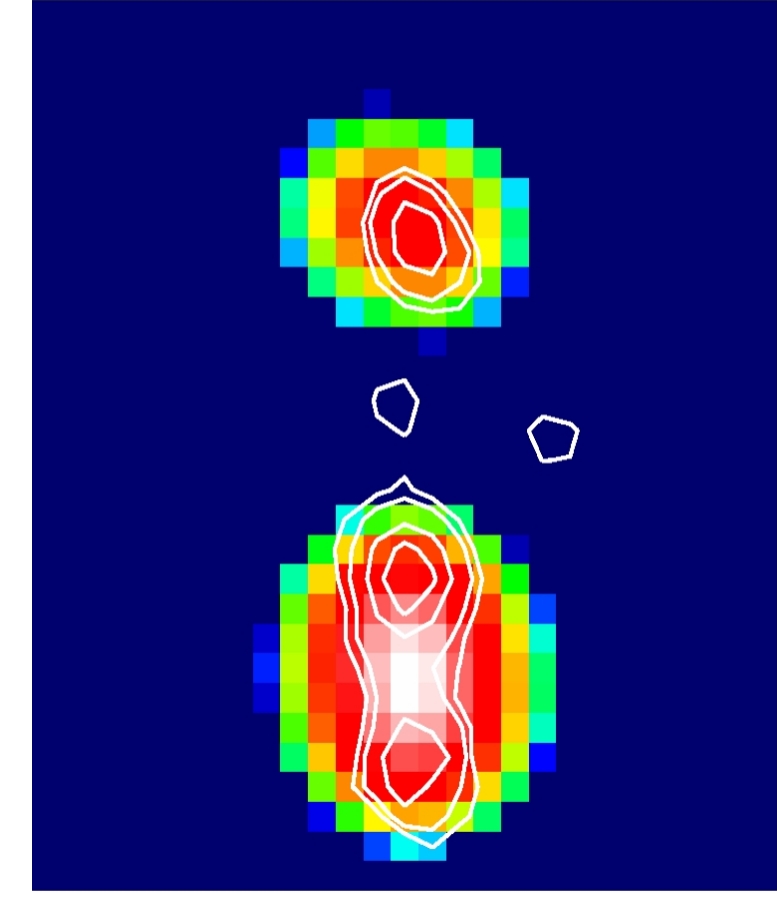}
\includegraphics[angle=0,width=40mm]{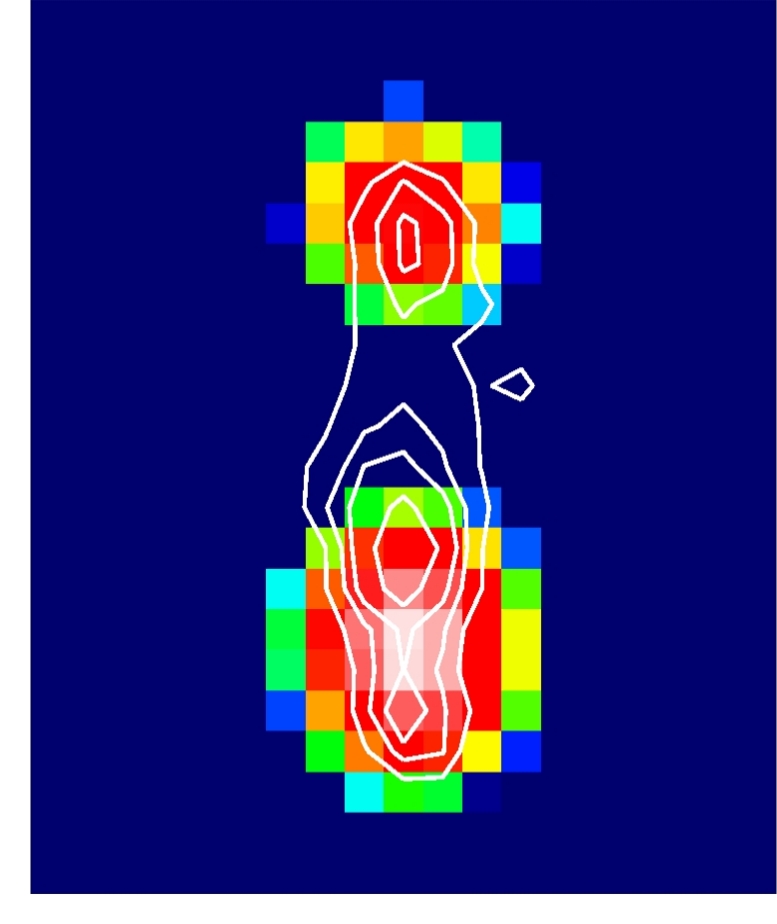}
\end{center}
\caption{Upper row: Observations of G030.50+00.95 taken at (left--right) \textit{Spitzer} 8\,$\mu$m (with 250\,$\mu$m contours), 
PACS 70 \& 160\,$\mu$m and SPIRE 250, 350 and 500\,$\mu$m. 
Lower row: Model output for G030.50+00.95 at (left--right) 8 and 70\,$\mu$m (example background only --- showing core in absorption 
in middle) and 160, 250, 350 and 500\,$\mu$m (in emission). Overlaid contours correspond to the observed data.} 
\label{model3095}
\end{sidewaysfigure*}

\begin{sidewaysfigure*}
\begin{center}
\includegraphics[angle=0,width=40mm]{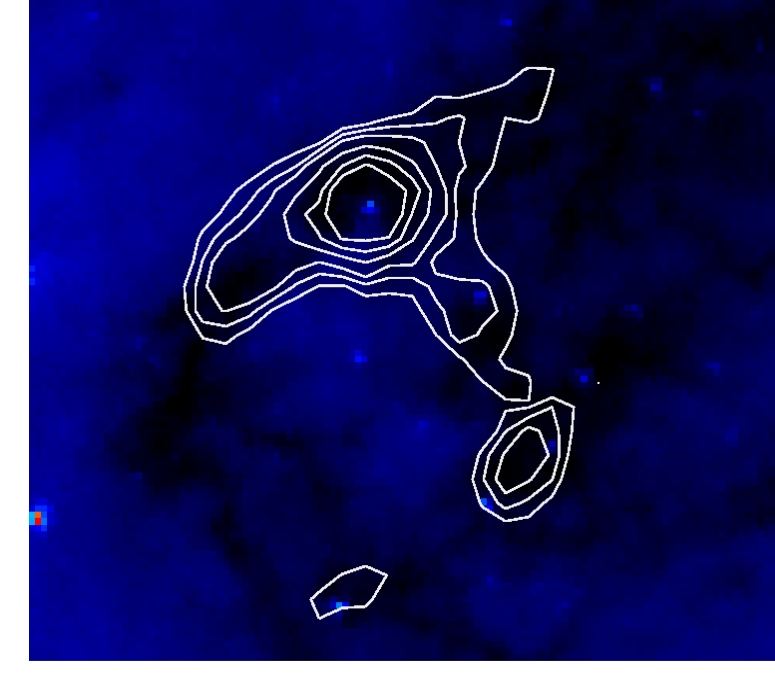}
\includegraphics[angle=0,width=40mm]{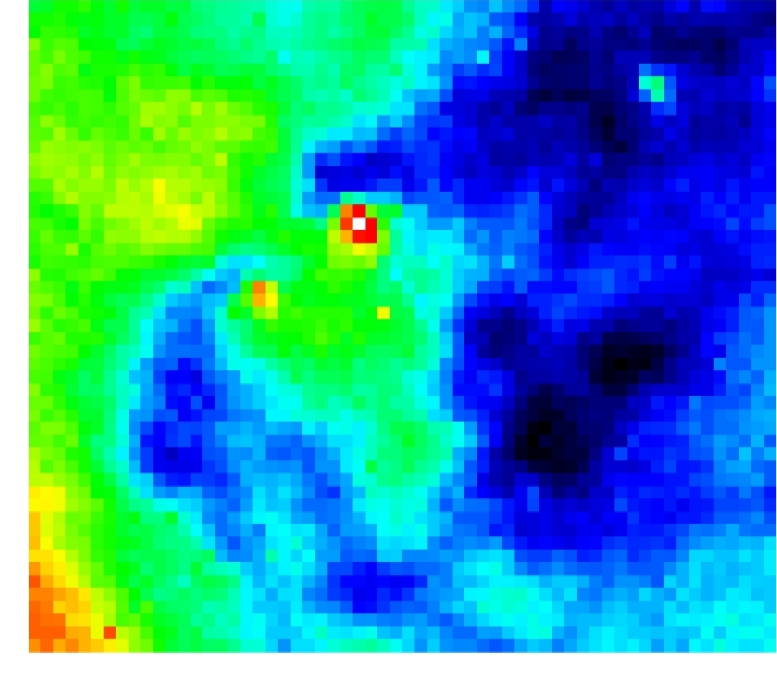}
\includegraphics[angle=0,width=40mm]{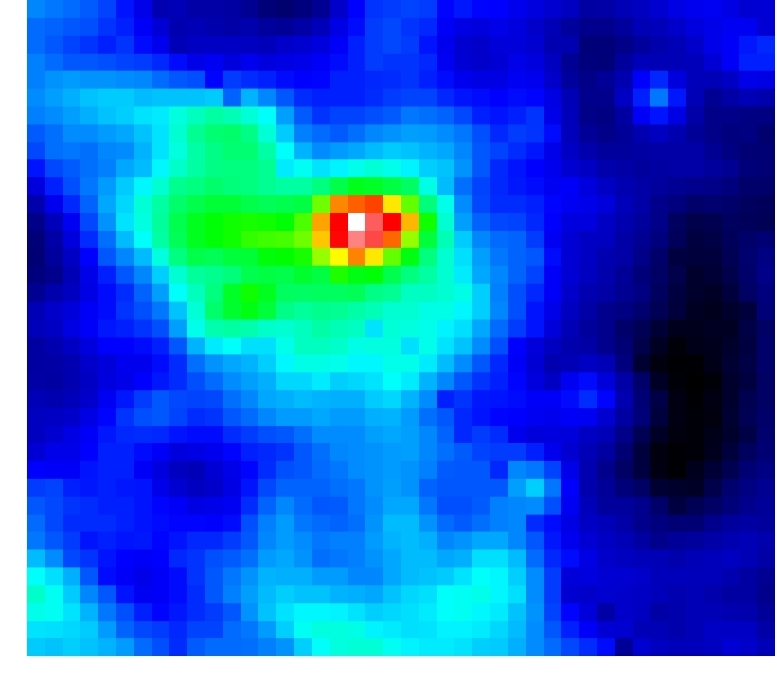}
\includegraphics[angle=0,width=40mm]{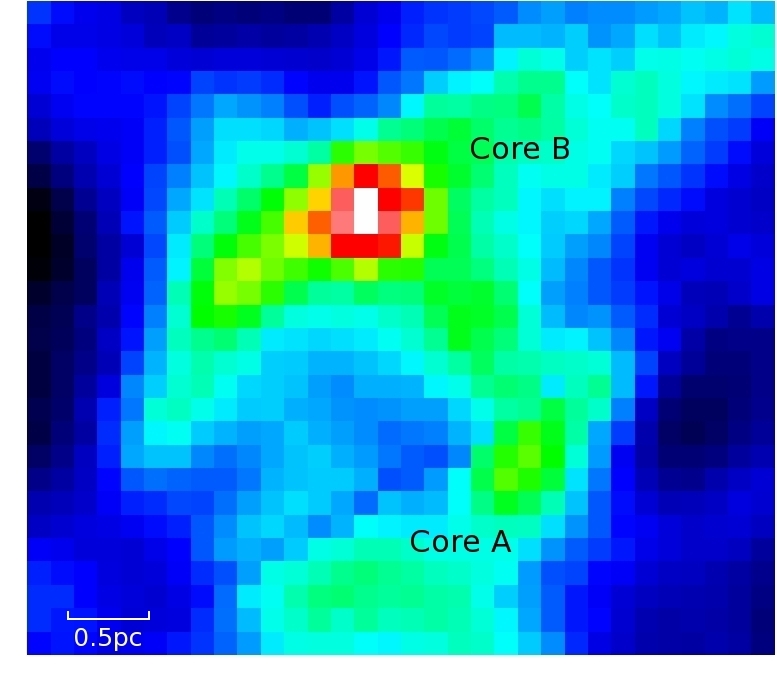}
\includegraphics[angle=0,width=40mm]{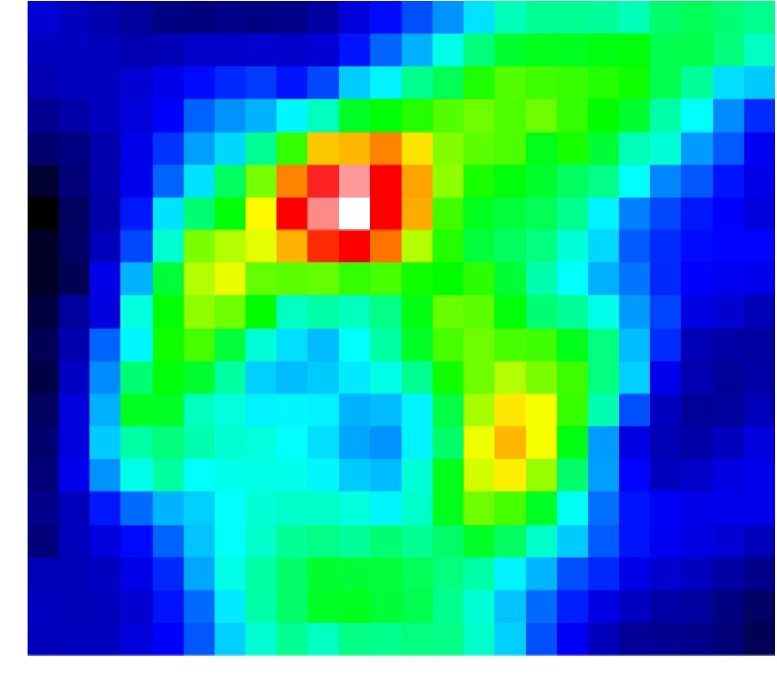}
\includegraphics[angle=0,width=40mm]{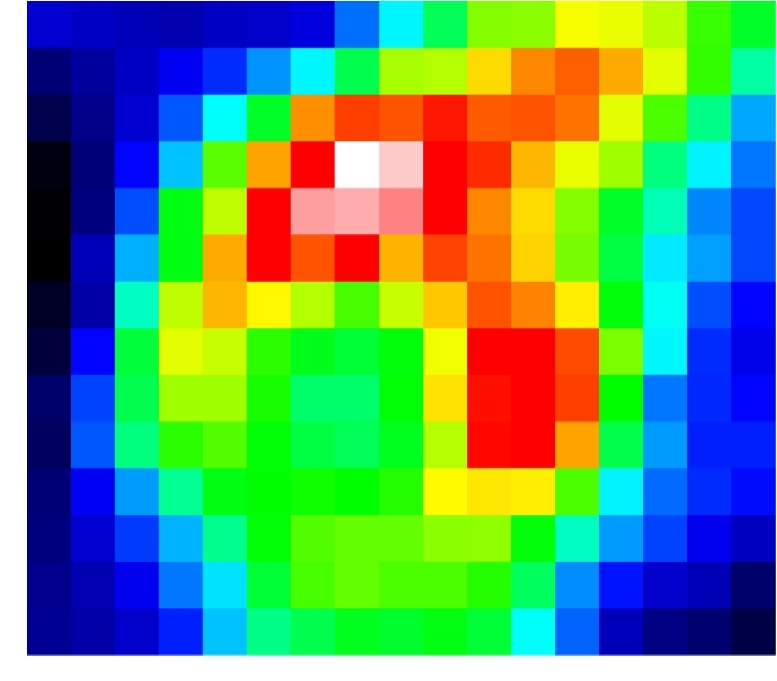}
\includegraphics[angle=0,width=40mm]{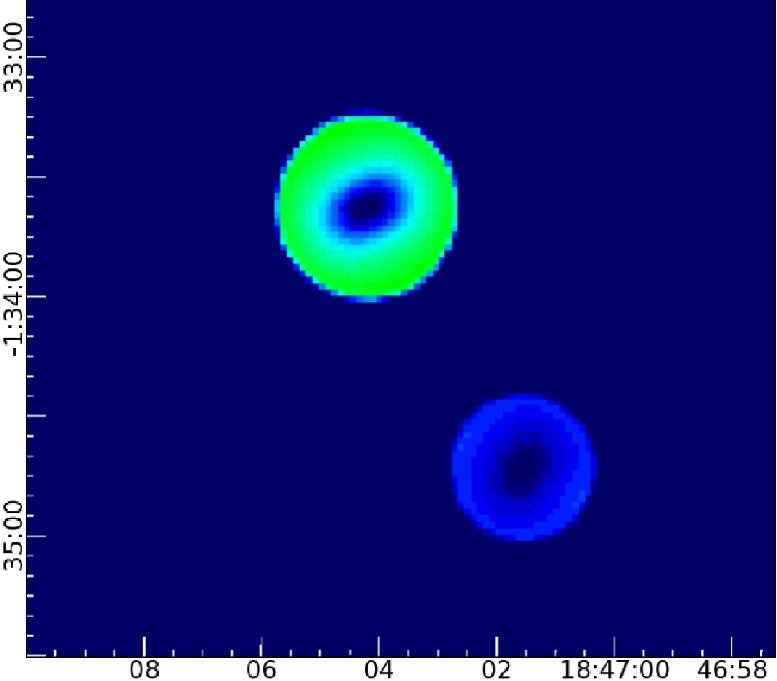}
\includegraphics[angle=0,width=40mm]{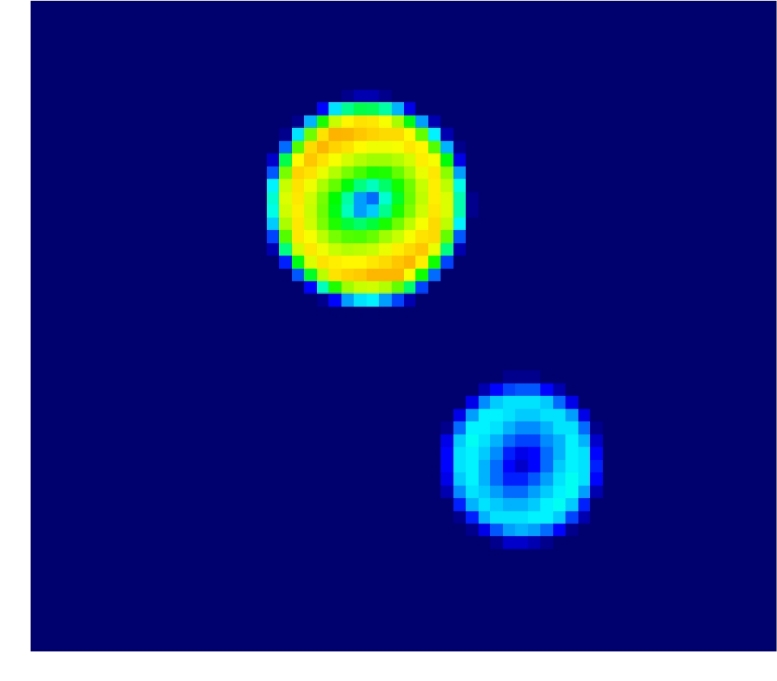}
\includegraphics[angle=0,width=40mm]{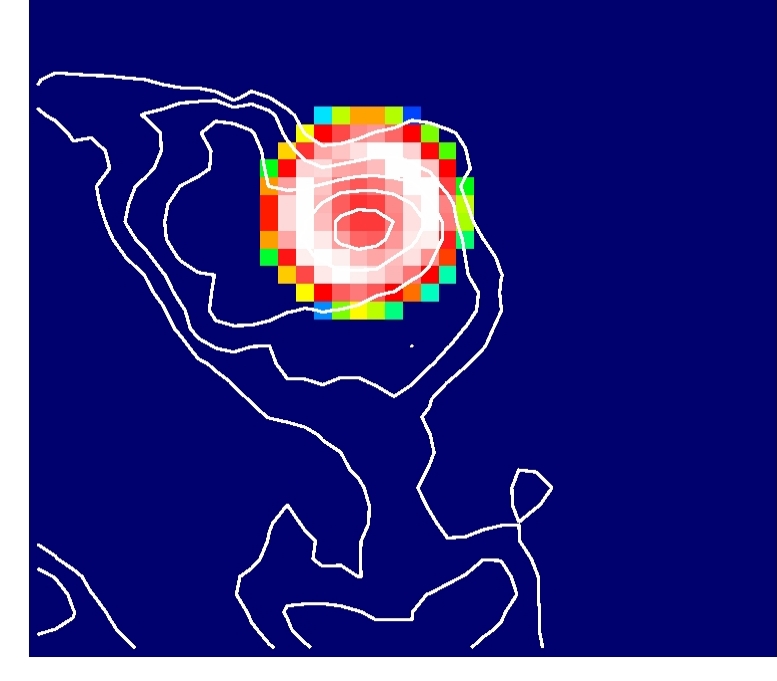}
\includegraphics[angle=0,width=40mm]{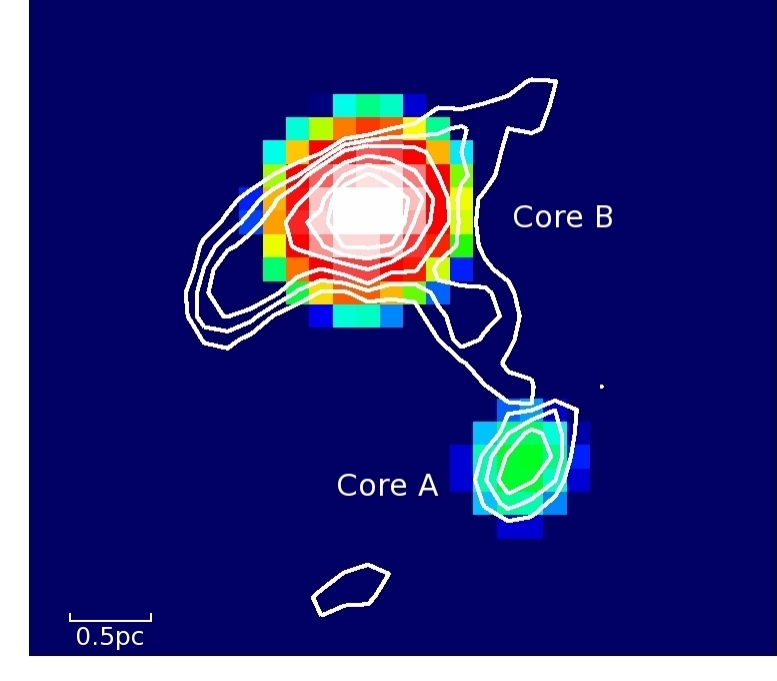}
\includegraphics[angle=0,width=40mm]{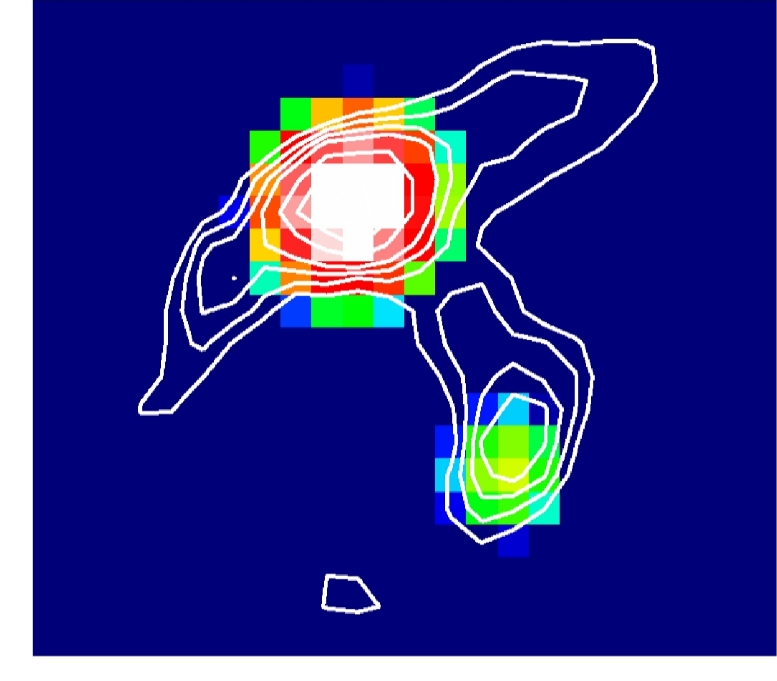}
\includegraphics[angle=0,width=40mm]{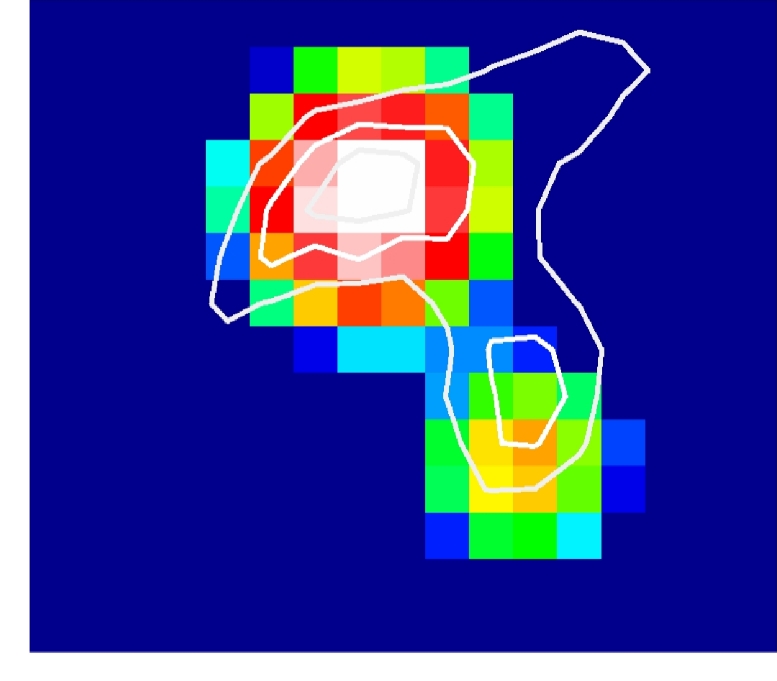}
\end{center}
\caption{Upper row: Observations of G031.03+00.26 taken at (left--right) \textit{Spitzer} 8\,$\mu$m (with 250\,$\mu$m contours), 
PACS 70 \& 160\,$\mu$m and SPIRE 250, 350 and 500\,$\mu$m. 
Lower row: Model output for G031.03+00.26 at (left--right) 8 and 70\,$\mu$m (example background only --- showing core in absorption 
in middle) and 160, 250, 350 and 500\,$\mu$m (in emission). Overlaid contours correspond to the observed data.} 
\label{model3126}
\end{sidewaysfigure*}

\begin{sidewaysfigure*} 
\begin{center}
\includegraphics[angle=0,width=40mm]{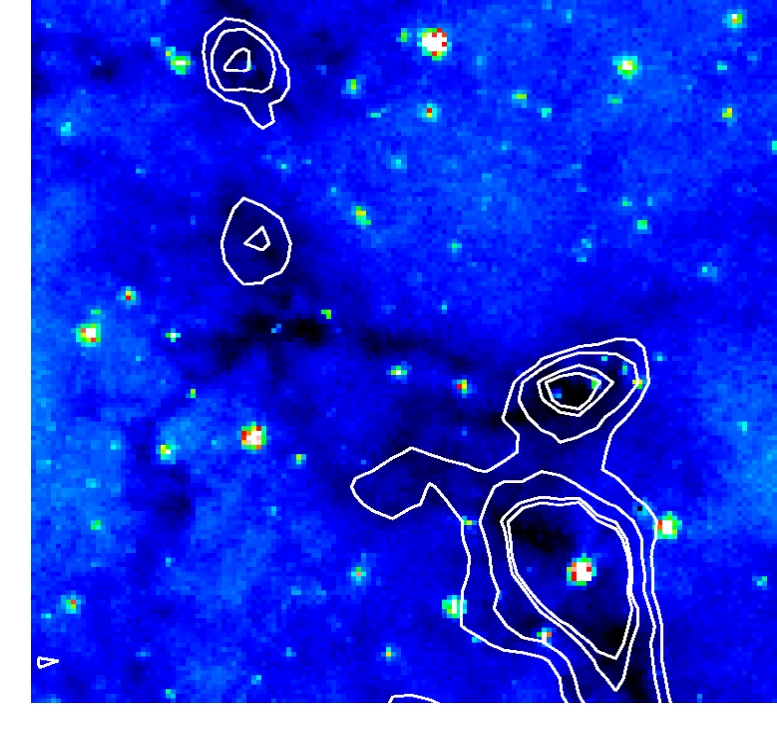}
\includegraphics[angle=0,width=40mm]{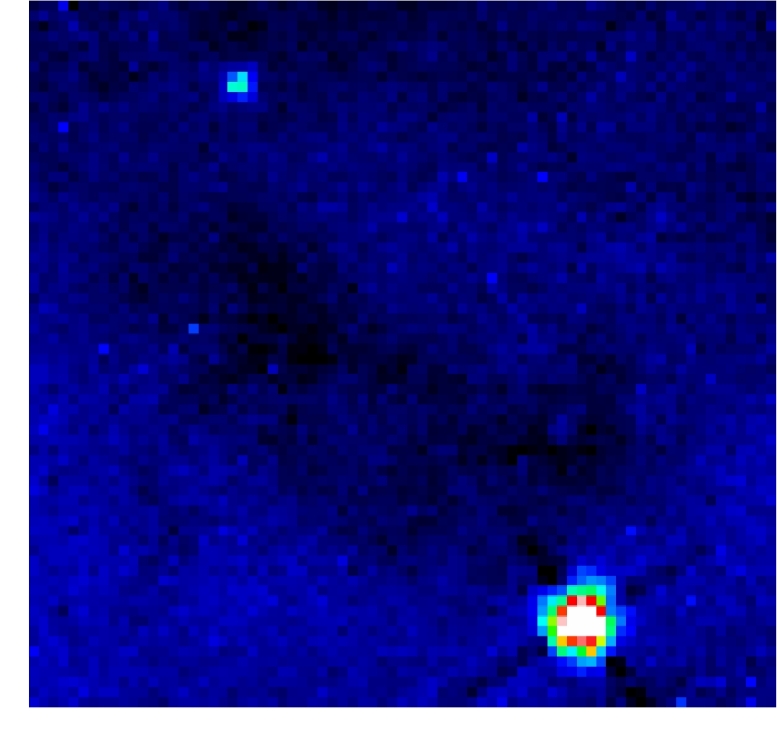}
\includegraphics[angle=0,width=40mm]{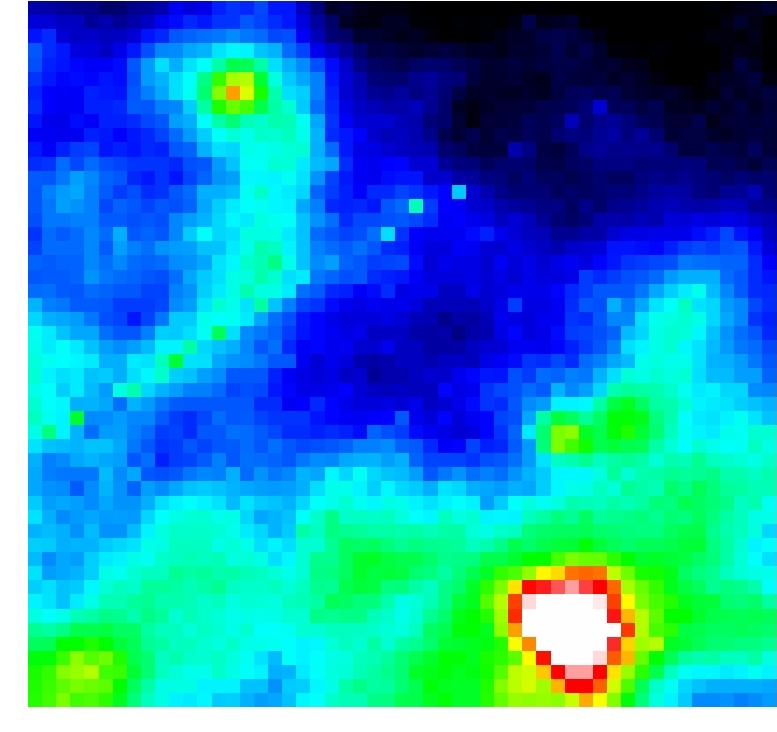}
\includegraphics[angle=0,width=40mm]{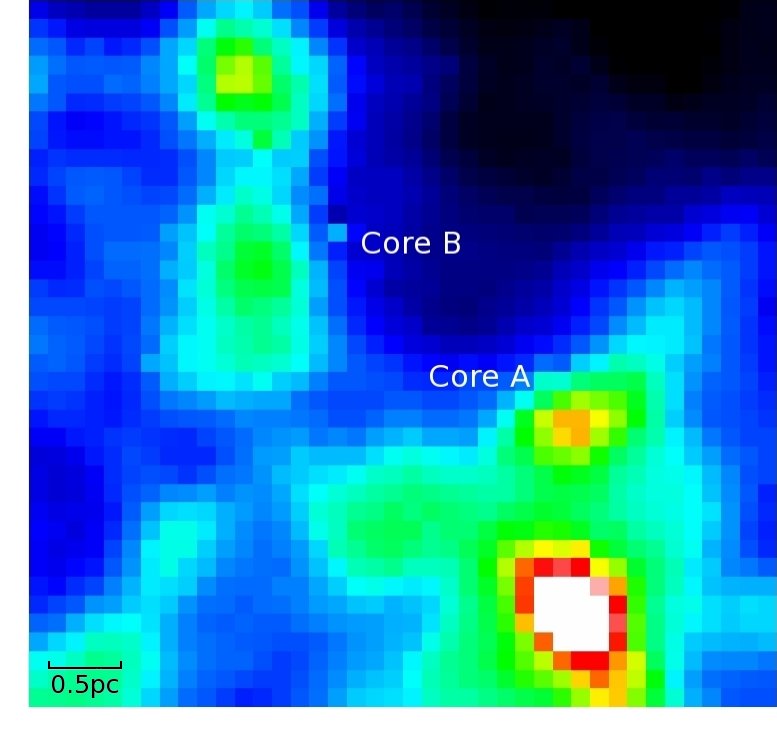}
\includegraphics[angle=0,width=40mm]{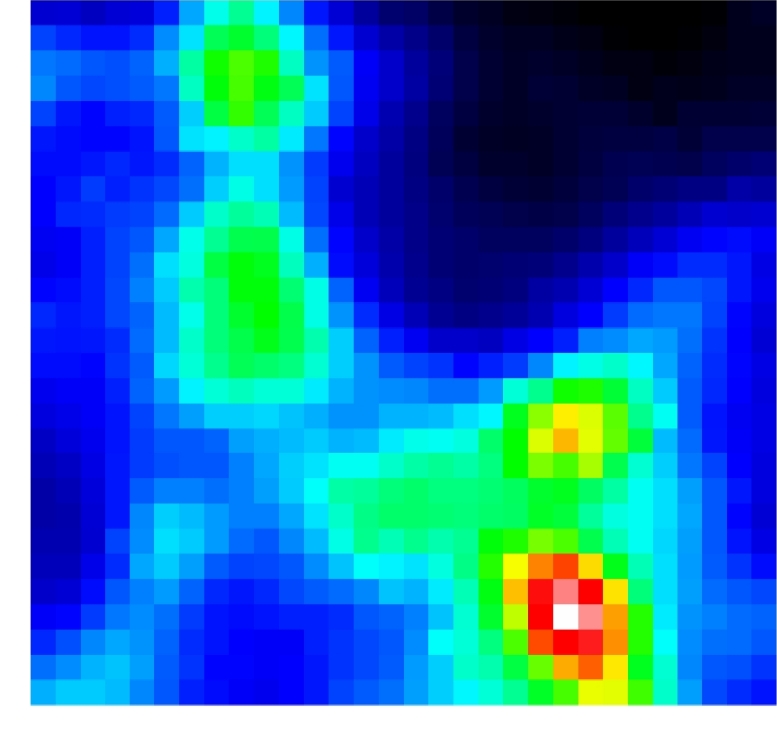}
\includegraphics[angle=0,width=40mm]{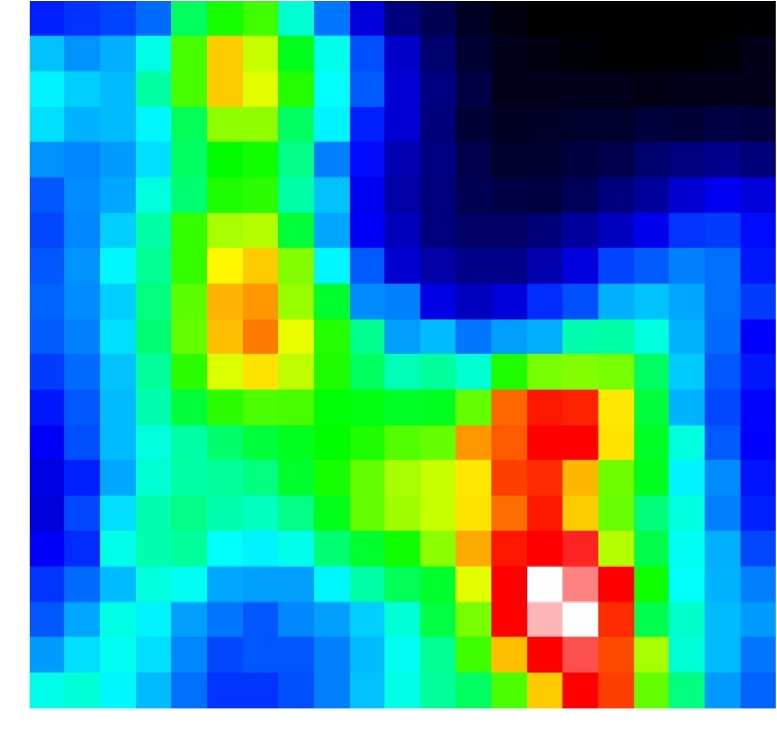}
\includegraphics[angle=0,width=40mm]{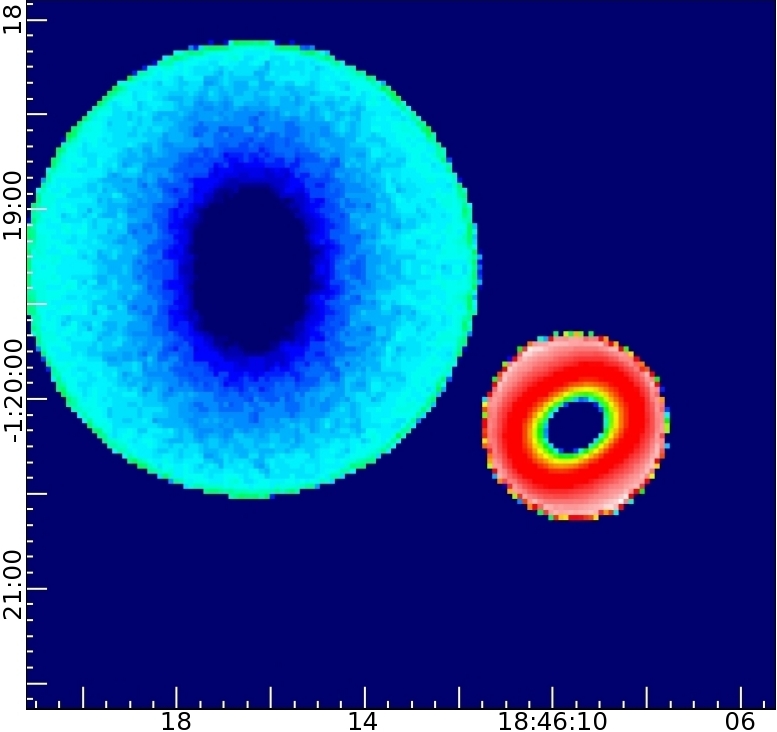}
\includegraphics[angle=0,width=40mm]{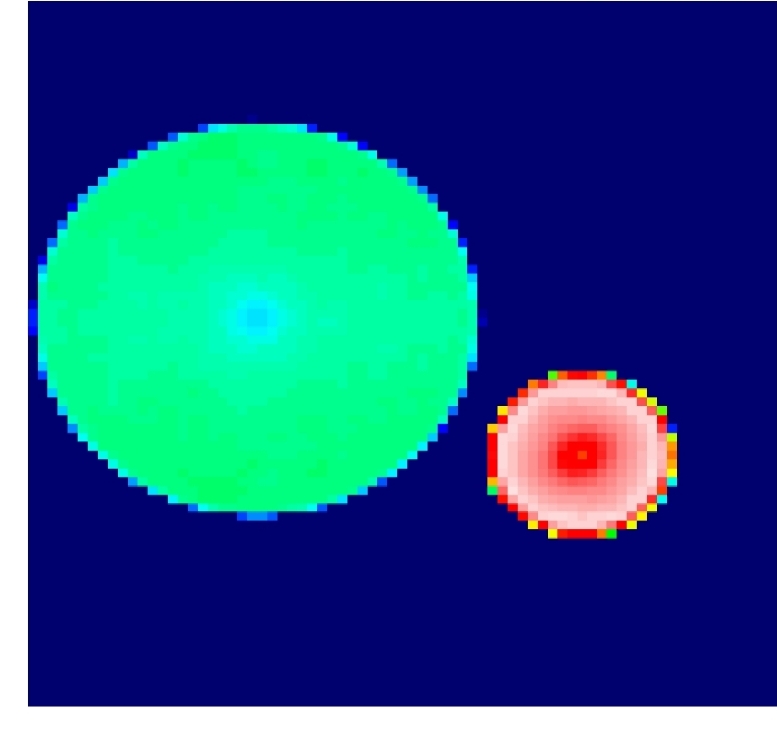}
\includegraphics[angle=0,width=40mm]{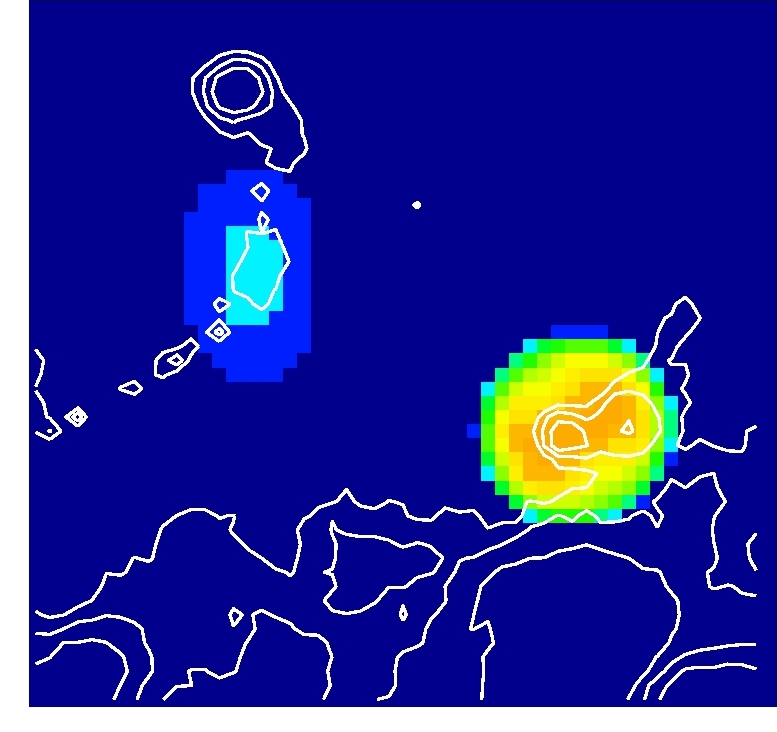}
\includegraphics[angle=0,width=40mm]{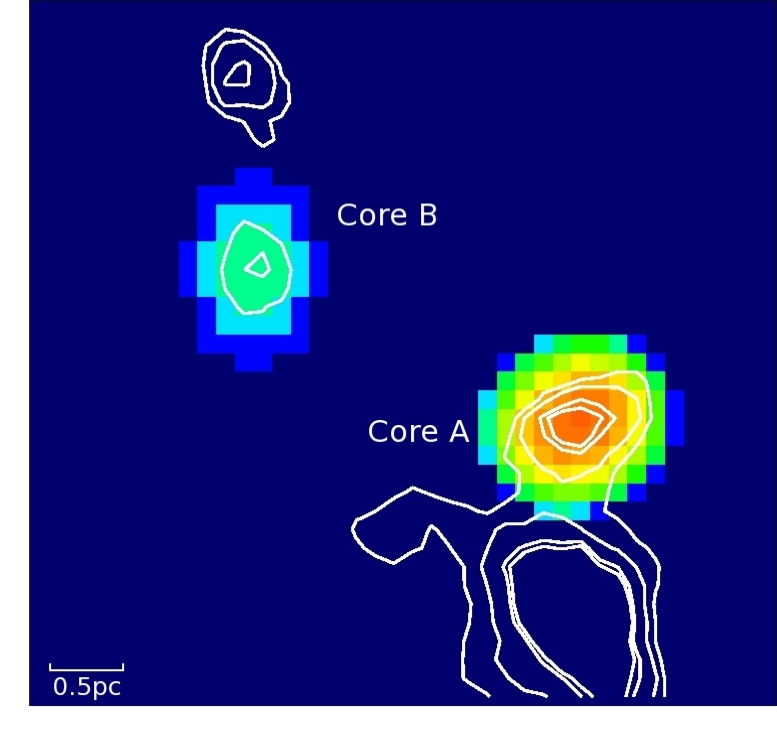}
\includegraphics[angle=0,width=40mm]{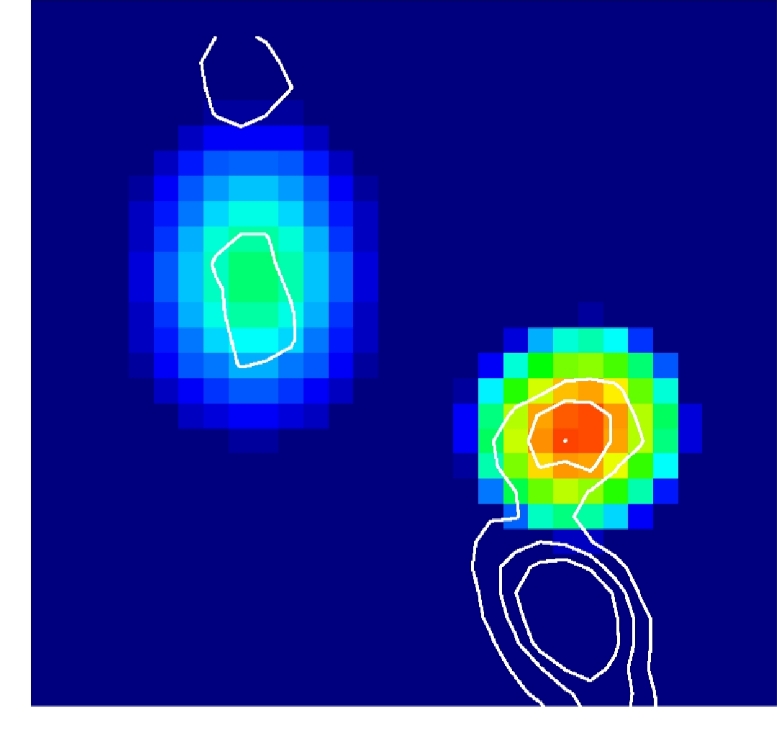}
\includegraphics[angle=0,width=40mm]{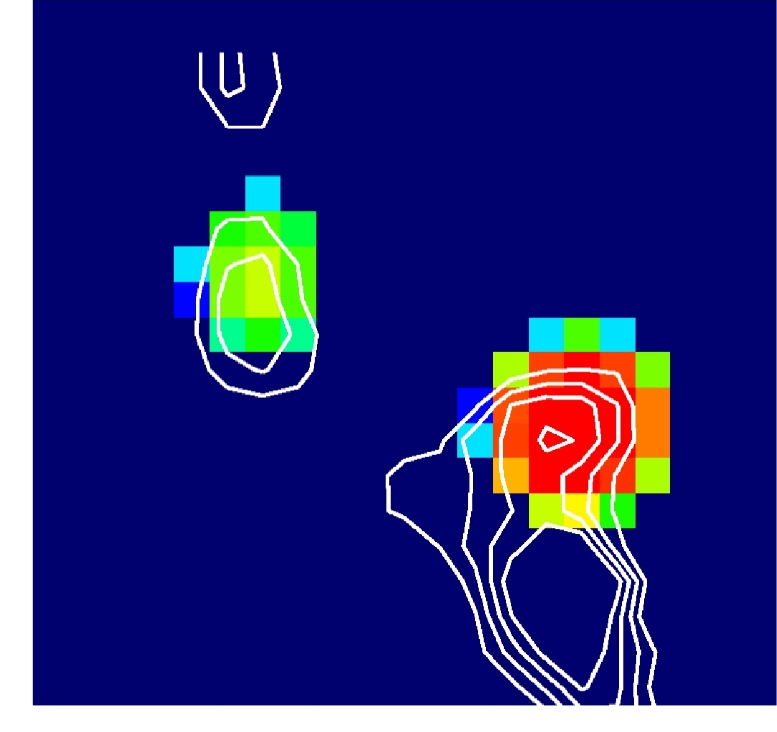}
\end{center}
\caption{Upper row: Observations of G031.03+00.76 taken at (left--right) \textit{Spitzer} 8\,$\mu$m (with 250\,$\mu$m contours), 
PACS 70 \& 160\,$\mu$m and SPIRE 250, 350 and 500\,$\mu$m. 
Lower row: Model output for G031.03+00.76 at (left--right) 8 and 70\,$\mu$m (example background only --- showing core in absorption 
in middle) and 160, 250, 350 and 500\,$\mu$m (in emission). Overlaid contours correspond to the observed data.} 
\label{model3176}
\end{sidewaysfigure*}

Our aim is to model those cores without embedded protostars. We focus on three objects, G030.50+00.95, 
G031.03+00.26 and G031.03+00.76, which meet the criteria for IRDCs. All three contain
two cores each, designated `Core A' and `Core B' in order of Right Ascension within each cloud. 
Positions and physical properties of all
six cores can be found in Table \ref{coreprop}, and images are displayed in Figs. \ref{model3095}--\ref{model3176}.

\begin{table*}
\begin{center}
\caption{The observed, integrated flux densities of the six cores.}
\label{flux}
\begin{tabular}{lllllll} \hline
Wavelength & \multicolumn{6}{c}{Flux Density (Jy)}\\
& \multicolumn{2}{l}{G030.50+00.95} & \multicolumn{2}{l}{G031.03+00.26} & \multicolumn{2}{l}{G031.03+00.76} \\
 & Core A & Core B & Core A & Core B & Core A & Core B \\ \hline 
70\,$\mu$m & $<$0.03 & $<$3.6 & $<$0.5 & $<$10.1 & $<$1.0 & $<$1.3\\
160\,$\mu$m & 13.4 & 41.9 & 10.9 & 60.3 & 13.7 & 15.4\\
250\,$\mu$m &  18.0 & 31.4 & 11.7 & 47.5 & 13.4 & 13.3\\
350\,$\mu$m & 9.6 & 16.0 & 6.8 & 23.0 & 8.8 & 8.8\\
500\,$\mu$m & 5.0 & 9.0 & 3.2 & 9.7 & 4.5 & 4.4\\
\hline
\end{tabular}
\end{center}
\end{table*}

\subsection{G030.50+00.95} 
In this cloud, Core B is the more extended and brighter of the two cores and located to the south of Core A 
(see Fig.~\ref{model3095}). 
In the 250\,$\mu$m and 350\,$\mu$m wavebands, it is possible to see that Core B can be separated into
two cores. However, as the individual peaks cannot be seen at either 160\,$\mu$m or 500\,$\mu$m, the two are grouped together and
modelled as a single core.  A distance of 3.4\,kpc for the entire cloud is assumed \citep{simon06b}. No previously calculated
masses are available for these cores. Results calculated from our data are presented in Table \ref{coreprop}.

\subsection{G031.03+00.26}
S06 find three cores in this cloud. However, only two can be seen as significant emission sources in the FIR and so
here the third is ignored. Core B is the more extended and brighter of the two and is located to the north of Core A (see Fig.
\ref{model3126}). Emission from Core B can be clearly seen at wavelengths as short as 160\,$\mu$m, 
whereas emission from Core A is not seen shortward of 250\,$\mu$m.

We take a distance of 4.9\,kpc for both cores \citep{teyssier02} although we note that there are two emission line components 
among the line of sight, with the second emission line giving a distance of 6.6\,kpc. 
Therefore there is some uncertainty in this distance. \citet{parsons09} use 850\,$\mu$m data and find
masses of 420 and 560\,M$_{\odot}$, respectively, for these cores. However, these authors used a different size of aperture. 
We make a direct comparison with these results in Section \ref{mass} below. 
Results calculated from our data are presented in Table \ref{coreprop}.

\subsection{G031.03+00.76}
Here, Core B is located to the north of Core A (Fig. \ref{model3176}) and, although it is the more extended of the two, it is 
actually fainter than Core A. Neither core can be seen in emission at wavelengths shorter than 160\,$\mu$m. 
A distance of 3.4\,kpc is assumed \citep{simon06b}. No previously calculated masses are available for these cores.

A third core in this cloud, located to the south of Core A, is not modelled here due to a bright 8$\mu$m source at the same 
position, implying the core may have an internal heating source. A possible fourth core exists to the north of Core B, we 
ignore this as S06 state that it belongs to a different IRDC with no known distance. Results calculated from our data 
are presented in Table \ref{coreprop}.

\section{IRDC Models}

We model the cores in two different ways: first, using a greybody single-temperature fit, and second, using the 3D Monte Carlo 
radiative transfer code, \textsc{Phaethon} \citep{stamatellos03, stamatellos05, stamatellos10}.

\subsection{Single Temperature Fitting of SEDs}

\begin{figure*}
\begin{center}
\includegraphics[angle=0,width=60mm]{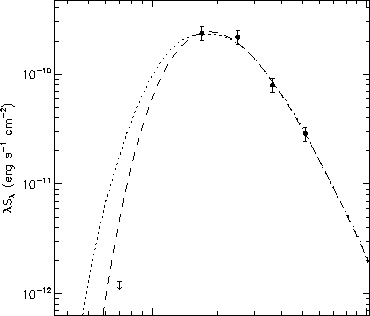}
\includegraphics[angle=0,width=60mm]{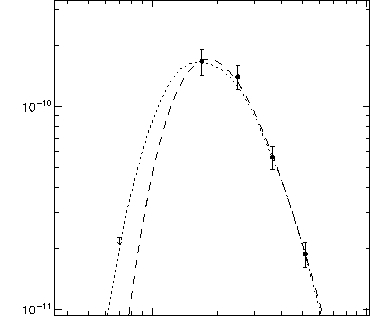}
\includegraphics[angle=0,width=60mm]{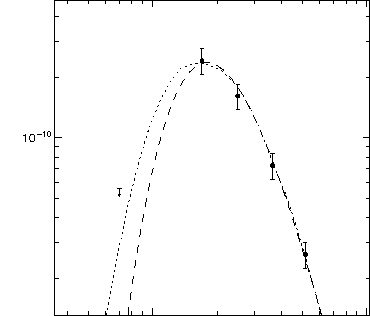}
\includegraphics[angle=0,width=60mm]{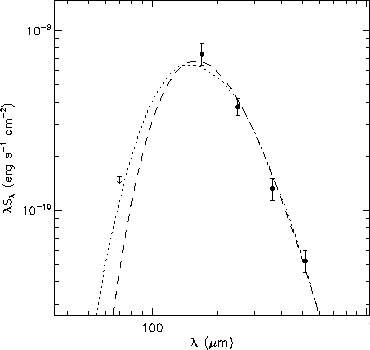}
\includegraphics[angle=0,width=60mm]{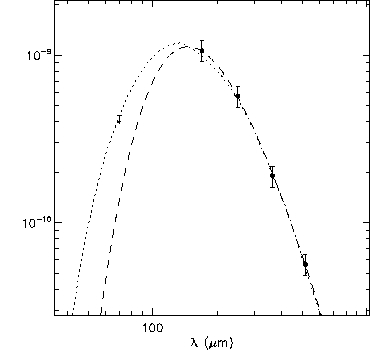}
\includegraphics[angle=0,width=60mm]{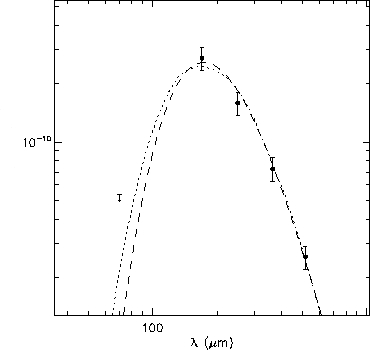}
\caption{SEDs of all six cores. Upper row (left--right): G030.50+00.95, G031.03+00.26 and G031.03+00.76 Core A. Lower row
(left--right): G030.50+00.95, G031.03+00.26 and G031.03+00.76 Core B. The dotted line shows the SED calculated from
\textsc{Phaethon}, using $\beta$=1.85. The dashed line shows the single-temperature grey-body, with temperatures as noted in 
Table \ref{coreprop}, also using $\beta$=1.85.}
\label{seda}
\end{center}
\end{figure*}

The flux density, integrated over twice the FWHM of each core using an elliptical aperture, was measured in all 
five FIR maps (see Table \ref{flux}) and an SED (spectral energy distribution) 
was plotted. These flux densities have been background subtracted, where the background was defined using an
off-cloud, elliptical aperture. For example, with G031.03+00.76 the area to the right of Core B was used. 
The background subtraction removed up to 50\% of the original flux, the exception being at 70\,$\mu$m 
where over 90\% of the original flux was removed. A single-temperature greybody (modelling thermal emission from cold 
dust) was fitted (using MPFit; \citealp{sedfit}) to each core, shown as a dashed line in Fig. \ref{seda}. 
This has the form
\begin{equation}
F_{\nu}=B_{\nu}(T) \, \Omega \, [\frac{2 m_H}{X} N(H_2) \kappa_\nu] ,
\end{equation}
where: $F_{\nu}$ is the flux density at frequency $\nu$; $B_{\nu}(T)$ is the blackbody function at temperature $T$; $\Omega$ is the
solid angle subtended at the observer by the source; 2$m_H$ is the mass of a hydrogen molecule; 
\textit{X} is the mass fraction of hydrogen and $\kappa_\nu$ is the dust mass opacity (e.g. \citealt{kirk10, wardthompson10}). 
$\kappa_\nu$ is given by
\begin{equation}
\kappa_{\nu}=0.1{\rm \,cm^{-2}g^{-1}}\times(\nu/1000{\rm GHz})^{\beta},
\end{equation}
where $\beta$, the dust emissivity index, was set to 1.85 \citep{ossenkopf94}. 

The model was fitted between 160\,$\mu$m and 500\,$\mu$m, with the flux density at 70\,$\mu$m
being used as an upper limit, as the cores are not visible at this wavelength. The temperature was allowed to vary over a range
3--40\,K. The best fit temperatures for the cores are given in Table \ref{coreprop}.

\subsection{Radiative Transfer Modelling of the Cores}
The cores were modelled using \textsc{Phaethon}, a 3D Monte Carlo radiative
transfer code. The code uses luminosity packets to represent the ambient
radiation field in the system. These packets are injected into the system where they interact (are absorbed, re-emitted or
scattered) with it stochastically. The ambient radiation field is taken to be a multiple of a modified version of the \citet{black94} 
interstellar radiation field, which gives a good approximation to the radiation field in the solar neighbourhood. 

The input variables of the code are the strength of the ambient radiation field, the density profile, the size and geometry 
of the core (i.e. spherical, flattened or cometary) and the dust properties of the system. The code calculates the
temperature profile of the system as well as SEDs and intensity maps, at different wavelengths and viewing angles.

All six cores showed some measure of eccentricity in the observations and so were modelled with a flattened geometry --- see
\citet{stamatellos04, stamatellos10} for details. In this case
the density profile is given by:
\begin{equation}
n \left( r, \theta \right) = n_0 \left( {\rm H}_2 \right) \frac{1 + A \left( \frac{r}{R_0} \right) ^2 \left[ \sin (\theta) 
\right] ^p }{\left[ 1 + \left( \frac{r}{R_0} \right) ^2 \right] ^2} ,
\end{equation}
where \textit{r} is the radial distance, $\theta$ is the polar angle and $R_0$ is the flattening radius (i.e. the radial distance 
for which the central density is approximately constant). \textit{n}$_0$(H$_2$) is the central density, which is controlled as an input 
variable. \textit{A} is a factor that controls the equatorial to polar optical depth ratio and determines how flattened the core is. 
\textit{p} determines how quickly the optical depth changes from equator to pole, and was set to 2.

The FWHM of the major axis of each observed core was measured at 250 $\mu$m and used as the model core's semi-major axis. The 
flattening radius, \textit{R$_0$}, was set at one tenth of this value. The central density, \textit{n}$_0$(H$_2$), and 
interstellar radiation field (ISRF) incident 
on the core were then varied until the output model's SED matched the observed data. The model SED is shown as a dotted line 
in Fig. \ref{seda}. The final values for \textit{n}$_0$(H$_2$) and the incident ISRF are noted in Table \ref{coreprop}.

It should be noted that the internal structure of cores within infrared dark clouds have not yet been observed in detail. 
Hence our assumption of an elliptical geometry may be an over-simplification. If there is structure on smaller scales than we can resolve this
might affect our results. For example, small scale fragmentation might allow the ISRF to penetrate further into the cores, meaning that our
calculated values are probably upper values.

Figs. \ref{model3095}--\ref{model3176} show the output of the model at wavelengths corresponding to the wavelengths of the observed 
data. The modelled images have pixels of 0.02$\times$0.02\,pc in size, which corresponds to 8\arcsec\ pixels for G031.03+00.26 and 14\arcsec{} 
pixels for G030.50+00.95 and G031.03+00.76. All the images have been convolved with the telescope beam. For wavelengths where no 
emission is visible in the model, an image showing background radiation is shown. As no attempt was made to correctly model the 
surrounding area, these images are not an accurate representation of the parent cloud but do show the cores in absorption as seen 
in the observations. 

With the possible exception of G31.03+00.26 Core B (which may have evidence of an 8\,$\mu$m point source), the 8\,$\mu$m 
images shows no emission in either 
the model or the observed data, as IRDCs do not emit significantly in the MIR due to their low temperatures. The modelled images show 
emission starting at 250\,$\mu$m for G031.03+00.26 Core A and 160\,$\mu$m for the remaining five cores (albeit only faintly 
for the cores in G031.03+00.76) and absorption at shorter wavelengths. The modelled emission fits well with the observations, 
at all wavelengths.

\begin{figure*}
\begin{center}
\includegraphics[angle=-90,width=90mm]{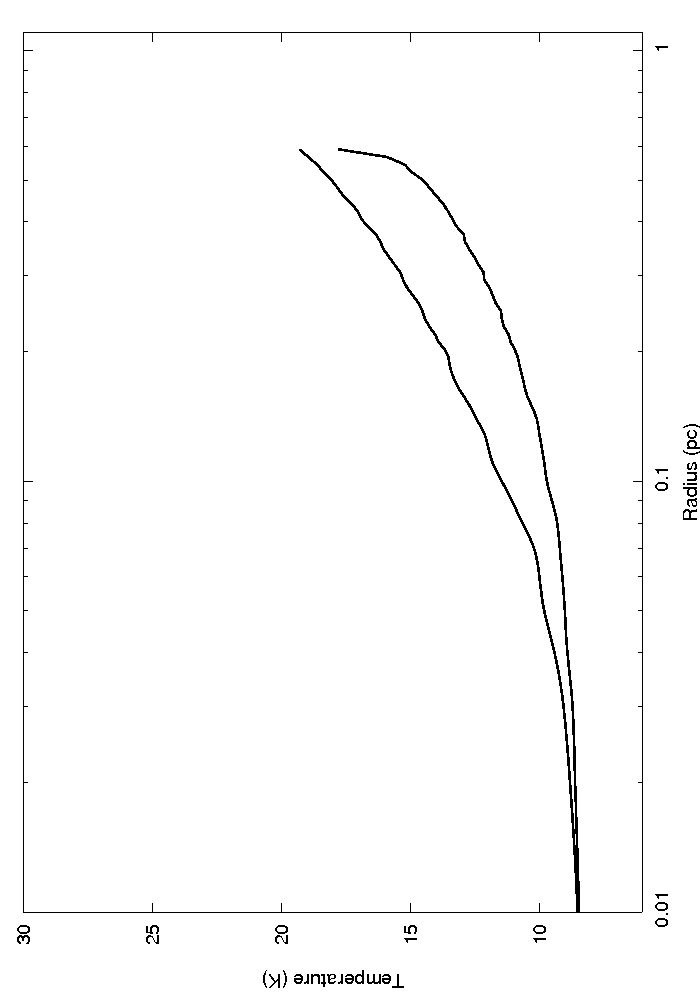}
\includegraphics[angle=-90,width=90mm]{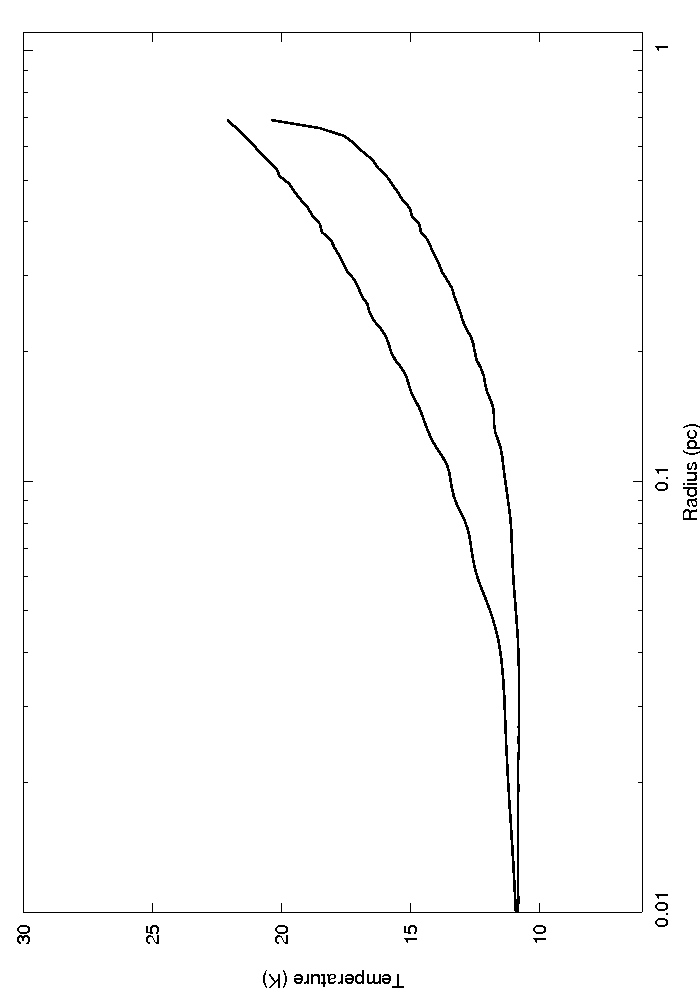}
\includegraphics[angle=-90,width=90mm]{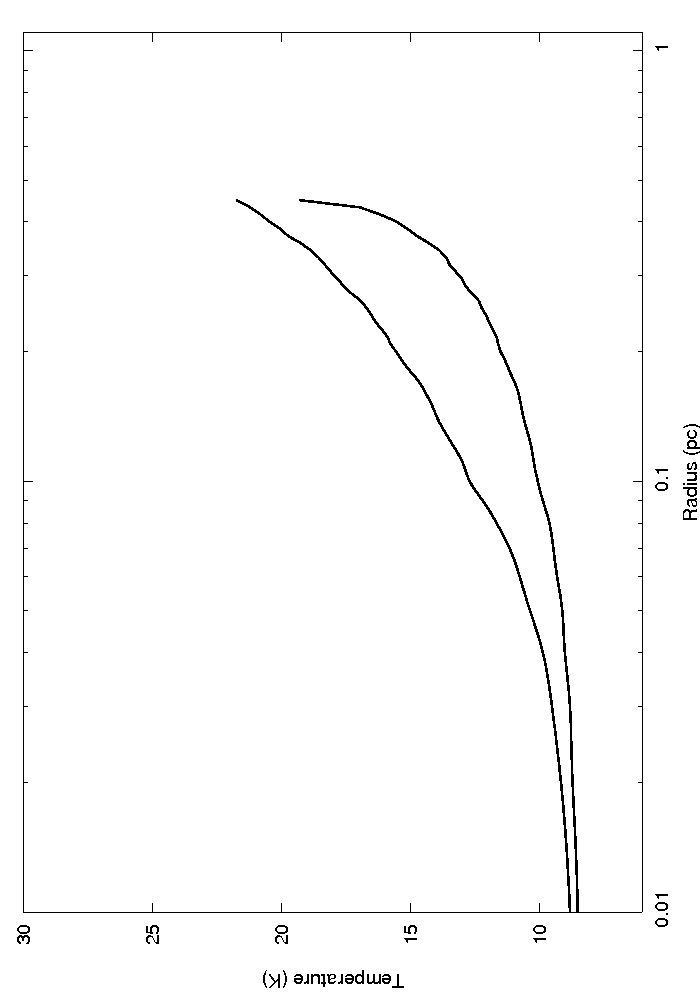}
\includegraphics[angle=-90,width=90mm]{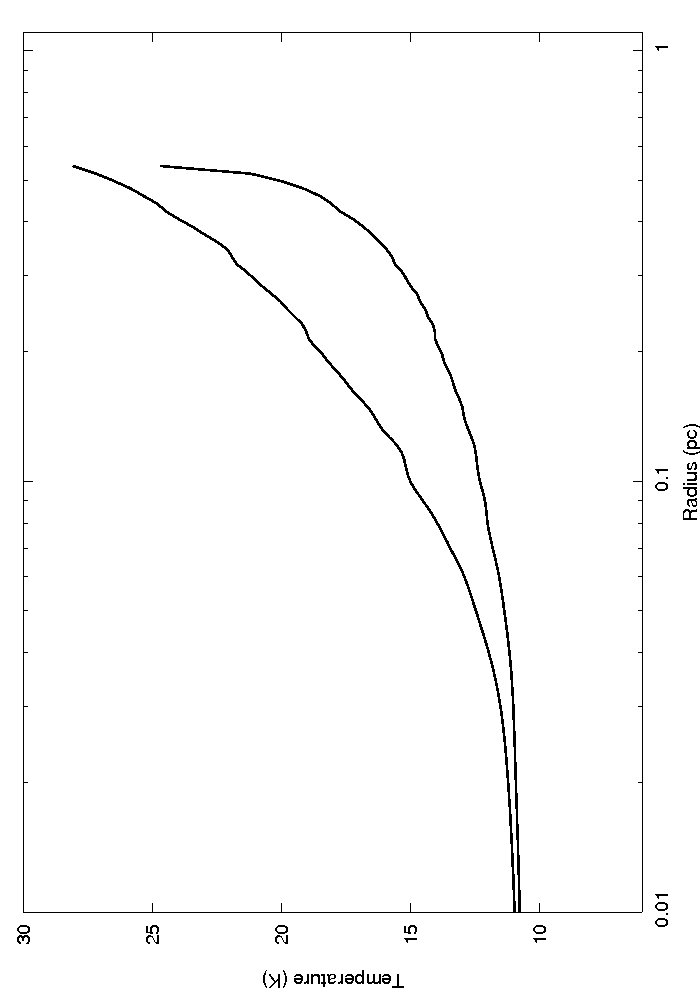}
\includegraphics[angle=-90,width=90mm]{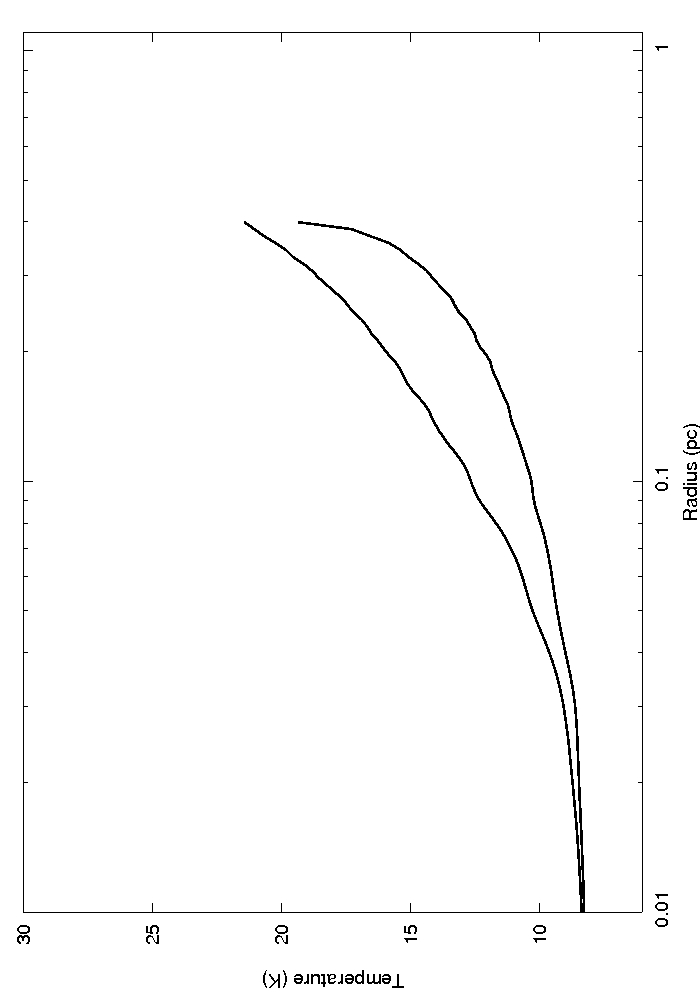}
\includegraphics[angle=-90,width=90mm]{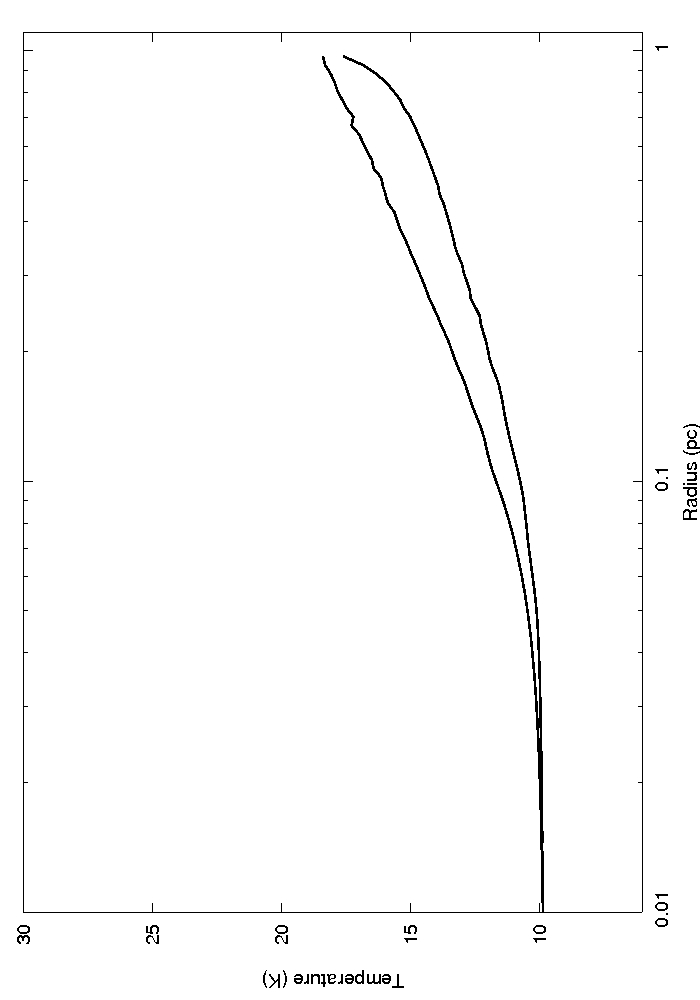}
\caption{Dust temperature profiles of the cores at two different directions of the cloud, as calculated from the radiative
transfer model. The lower line corresponds to the midplane of the flattened structure ($\theta=90\degr$), whereas the upper line
corresponds to the direction perpendicular to the midplane ($\theta=0\degr$). Upper row: G030.50+00.95 Cores A and B; Middle row:
G31.03+00.26 Cores A and B; Lower row: G31.03+00.76 Cores A and B}
\label{temprad}
\end{center}
\end{figure*}

\section{Discussion}
\subsection{Core Masses}
\label{mass}
The masses of the cores were calculated in two ways: (i) using the radiative transfer models; and (ii) using the observed flux
densities at 500\,$\mu$m. 
The latter was carried out using the equation 
\begin{equation}
M=\frac{F_{\nu}D^{2}}{B_{\nu}(T)\kappa_{\nu}},
\end{equation} 
\citep{hildebrand83}, where \textit{D} is the source distance. We call the former the model mass and the latter the SED mass.

Two different values of opacity at 500\,$\mu$m were used: $\kappa$=0.03\,cm$^{2}$\,g$^{-1}$ \citep{ossenkopf94} and 
$\kappa$=0.05\,cm$^{2}$\,g$^{-1}$ \citep{preibisch93, henning95}. The latter is the value used
to calculate the masses of low-mass prestellar cores (e.g. \citealt{kirk05}). The model masses were found using the
\citet{ossenkopf94} opacities.

The model masses match well with the SED masses calculated using $\kappa$=0.05\,cm$^{2}$\,g$^{-1}$, but not with those calculated
using $\kappa$=0.03\,cm$^{2}$\,g$^{-1}$.
Both sets of SED masses and the model masses are shown in Table \ref{coreprop}. 

The masses calculated here are in agreement with those found in previous studies of IRDCs 
(e.g. \citealt{williams01, simon06b, rathborne06}). In particular, in \citet{rathborne06}, masses of dark cores were calculated and
found to vary from 10 to 2100M${_\odot}$. All of our cores fit within this range.

850\,$\mu$m masses for both cores in G031.03+00.26 were calculated by \citet{parsons09}. Core A was found to have a mass of 
460\,M$_{\odot}$ and Core B of 520\,M$_{\odot}$. When using the same core boundaries as in \citet{parsons09} and 
$\kappa$=0.05\,cm$^{2}$\,g$^{-1}$, we find 500\,$\mu$m masses of 470\,M$_{\odot}$ and 540\,M$_{\odot}$, similar to the 
850\,$\mu$m masses. This gives us additional confidence in our calculations.

\subsection{Core Temperatures} 

The energy deposited in the gas due to cosmic ray ionisation heating is much lower than the energy absorbed by the dust 
due to external heating from the ISRF \citep{evans01}. Thus, we can safely assume that the dust temperature is not
affected by the energy exchange between gas and dust. The lack of MIR point sources also implies that heating via 
compression is minimal \citep{battersby10}. These cores (again with the possible exception of G031.03+00.26 Core B) 
therefore have no significant internal heating sources and are heated mainly by the external radiation field. 

The temperature profiles of the modelled cores are shown in Fig \ref{temprad}. They show a low temperature 
throughout the centre of the core, before it eventually begins rising, peaking at the outer edges. A higher temperature at the edges 
than at the centre is expected due to the lack of any internal heating source. Some stepping can be seen in
the graphs, this is due to the temperature table having a spacing of 0.2\,K. 

This gradient is consistent with previous studies of IRDCs (e.g. \citealt{peretto10}) and is 
similar to what is seen in low-mass prestellar cores (e.g. \citealt{nutter08}). However, IRDCs have higher outer temperatures than
low-mass prestellar cores, which can be explained by a higher ISRF around these IRDCs relative to nearby regions of star formation.

As all six cores have a flattened geometry, the temperature-radius profile varies with the direction inside the core. 
The two extreme cases, $\theta=0\degr$ and $\theta=90\degr$, are shown in the temperature-radius profiles. $\theta=90\degr$ is along
the midplane of the structure, while $\theta=0\degr$ corresponds to the short axis of a flattened core, perpendicular to the midplane.

The average temperature from the model of each core agrees with the result from the single-temperature greybody fit. However, the
temperature-radius profile shows this fit to be an over-simplification, with temperatures within each core varying by $\sim$10\,K
and being dependent upon angular direction in the core.

\subsection{The Interstellar Radiation Field} \label{DRF}
Five of the six cores required a higher ISRF than the solar neighbourhood value in order to correctly 
match the observed flux densities at shorter wavelengths (i.e. 160 and 250\,$\mu$m).
As an example we show the cores' ISRF at far-ultraviolet (FUV) in Table \ref{coreprop}. This is merely to indicate how the ISRF varies 
from core to core. For comparison, the solar neighbourhood ISRF at 
FUV is 7.6$\times$10$^{-3}$\,erg~s$^{-1}$~cm$^{-2}$ \citep{black94} or 4.6 times the Habing flux 
(1.6$\times10^{-3}$\,erg~s$^{-1}$~cm$^{-2}$, \citealt{habing68}). The IRDCs typically have an ISRF 
around four times greater, consistent with them being in the stronger interstellar radiation field of the inner Galactic Plane.

Differences between the ISRF required for cores within the same field can be explained by one core being more deeply buried within 
the parent cloud and thus having less external radiation heating it. The difference between the ISRFs of the cores 
are listed in Table \ref{coreprop} as $ln\left(\frac{I_1}{I_2}\right)$, where $I_1$ corresponds to the core that is modelled using a higher ISRF 
and $I_2$ to the core modelled with the lower ISRF (within the same field). This ratio can be thought as a measure of how more 
deeply a core is embedded when compared with the other core in the same field.

Core B in G031.03+00.26 appears to reside in an unusually intense radiation field, 17 times that of the solar neighbourhood
and far higher than any of the other cores, possibly pointing to an abnormally high ISRF in this field. Core A in the same cloud
has an ISRF value comparable to the other cores, which implies that it is either more deeply buried within the IRDC than Core B or there is
a highly asymmetrical external radiation field incident on this cloud. Alternatively, differences in emission from the cores could 
be the result of different physical properties (e.g. dust grain size) between different cores in the same IRDC. Core B may not, 
in fact, be completely externally heated. This one is unusual in our sample, in that it may have an embedded protostar. There is 
some evidence for an 8\,$\mu$m source associated with this core.

\subsection{Internal Heating Sources} \label{heat}

A lack of significant internal heating in the modelled cores was assumed due to the absence of any near-infrared or MIR emission. 
This is further supported by the temperatures of the cores decreasing towards their centres. However, more detailed study into the 
heating of infrared dark cores needs to be performed before this can be conclusively proven. For now we can put limits on the
luminosity of any embedded source.

Using the column-density estimated for each core, we can calculate the visual extinction, the K-band extinction,
and finally the extinction at 8\,$\mu$m (\textit{A$_8$}; \citealt{indebetouw05}). A peak 8\,$\mu$m flux was measured for each core
($F_{8}$) and, using
\begin{equation}
F = F_8\, 10\,^{A_8/2.5},
\end{equation}
the maximum flux density of any embedded source (\textit{F}) was calculated. This was then converted to a luminosity (\textit{L}) 
by multiplying by
the \textit{Spitzer} 8\,$\mu$m bandwidth to convert from flux density to flux and using 
\begin{equation}
L=4 \pi D^2 F .
\end{equation}

These values are listed in Table \ref{coreprop}. We note that G31.03+00.76 Core A has some 8\,$\mu$m sources around it's edge, but they do
not appear to be centred on the core.

Our upper limits for G30.50+00.95 Cores A and B and G31.03+00.76 Core B are fairly low. We can thus rule out any significant 
massive star formation having occurred in their centres, making it possible that these cores are the high-mass equivalents of low-mass
prestellar cores. For both cores in G031.03+00.26 and in G31.03+00.76 Core A, the upper limits are too high to reach this conclusion.

In the case of G031.03+00.26 Core B a very faint source can be seen at 8\,$\mu$m from which $F_{8}$ was measured. These 
calculations make the assumption that this source is deeply embedded within the cloud. Should the source instead be in the 
foreground or towards the nearer edge of the IRDC then the maximum embedded luminosity would be very much lower.

\section{Summary and Conclusions}

We have confirmed the status of G030.50+00.95, G031.03+00.26 and G031.03+00.76 as IRDCs using mid- and far-infrared observations.
Six infrared dark cores have been found in the FIR in these IRDCs and modelled using the 3D Monte Carlo radiative transfer code
\textsc{Phaethon}. The radiative transfer models reproduce the observed SEDs and the FIR images of these cores. 

The \textsc{Phaethon} code calculated masses for the cores ranging from 90 to 290\,M$_{\odot}$. The masses calculated from the SED 
temperature ranged from 100\,M$_{\odot}$ to 310\,M$_{\odot}$ (with $\kappa$=0.05\,cm$^{2}$\,g$^{-1}$). These are in agreement with 
previous masses found for infrared dark cores. It should be noted that objects of this size and mass are not likely to create a single
 star, but rather a cluster of stars. 

The greybody fitted SED temperatures varied from 14 to 17\,K. The average temperatures calculated from the model agreed with these 
values. However, the model showed that the single-temperature fit was an over-simplification of the temperatures inside the cores as 
temperature-radius profiles showed changes in temperature of $\sim$10 to 17\,K within a single core. The model calculated temperatures
of 8 to 11\,K at the centre of each core and 18 to 28\,K at the surface. 

Significant temperature gradients were also seen by \citet{peretto10}, where temperature maps of 22 IRDCs were
created and non-uniform temperatures ranging from 10 to 22\,K were found. \citet{stamatellos10} used 
\textsc{Phaethon} to model an IRDC core and found temperatures from 10 to 21\,K, consistent with our results.

The maximum luminosity of a star embedded within each core was found. The upper luminosity limits for three of the cores 
(G30.50+00.95 Cores A and B and G31.03+00.76 Core B) are low, ruling out the possibility of any significant high mass star formation 
having yet occured and making it likely that these cores are the high-mass equivalent of prestellar cores.

In most cases, the amount of radiation incident on the cores was found to be higher than that in the local neighbourhood, as expected
for objects within active star-forming regions in the Galactic Plane. Differences in the amount of radiation falling on cores in 
the same cloud were attributed to one core being more deeply buried within the IRDC than the other. This method can now be used 
on all cores within IRDCs found in the Hi-GAL survey.

\section*{Acknowledgements}
LAW acknowledges STFC studentship funding.
SPIRE has been developed by a consortium of institutes led by Cardiff University (UK) and including Univ. 
Lethbridge (Canada); NAOC (China); CEA, LAM (France); IFSI, Univ. Padua (Italy); IAC (Spain); 
Stockholm Observatory (Sweden); Imperial College London, RAL, UCL-MSSL, UKATC, Univ. Sussex 
(UK); and Caltech, JPL, NHSC, Univ. Colorado (USA). This development has been supported by national 
funding agencies: CSA (Canada); NAOC (China); CEA, CNES, CNRS (France); ASI (Italy); MCINN 
(Spain); SNSB (Sweden); STFC (UK); and NASA (USA). 
PACS has been developed by a consortium of institutes led by MPE (Germany) and including UVIE
(Austria); KU Leuven, CSL, IMEC (Belgium); CEA, LAM (France); MPIA (Germany); INAF-IFSI/OAA/OAP/OAT, LENS, SISSA (Italy); IAC
(Spain). This development has been supported by the
funding agencies BMVIT (Austria), ESA-PRODEX (Belgium), CEA/CNES (France), DLR (Germany),
ASI/INAF (Italy), and CICYT/MCYT (Spain). 
HIPE is a joint development by the \textit{Herschel} Science Ground 
Segment Consortium, consisting of ESA, the NASA \textit{Herschel} Science Center, and the HIFI, PACS and 
SPIRE consortia.
This work is also based, in part, on observations made with the \textit{Spitzer} Space Telescope, which
is operated by the Jet Propulsion Laboratory, California Institute of Technology under a contract with NASA.

\bibliographystyle{aa}
\bibliography{bib}

\end{document}